\documentclass{article}%
\usepackage{amsmath}
\usepackage{graphicx}%
\usepackage{amsfonts}%
\usepackage{amssymb}
\newtheorem{theorem}{Theorem}

\newtheorem{corollary}[theorem]{Corollary}

\newtheorem{definition}[theorem]{Definition}
\newtheorem{example}[theorem]{Example}

\newtheorem{notation}[theorem]{Notation}

\newtheorem{remark}[theorem]{Remark}

\newenvironment{proof}[1][Proof]{\textbf{#1.} }{\ \rule{0.5em}{0.5em}}

\begin{document}

\title{Topics in asynchronous systems}
\author{Serban. E. Vlad\\Str Zimbrului, Nr.3, Bl.PB68, Ap.11, 410430, Oradea, Romania\\serban\_e\_vlad@yahoo.com, www.geocities.com/serban\_e\_vlad}
\date{}
\maketitle

\begin{abstract}
In the paper we define and characterize the asynchronous systems from the
point of view of their autonomy, determinism, order, non-anticipation, time
invariance, symmetry, stability and other important properties. The study is
inspired by the models of the asynchronous circuits.

\end{abstract}

\bigskip



\bigskip

\textbf{Contents }1.\textbf{ }Introduction, 2. Preliminaries, 3. Asynchronous
systems, 4. Initial states, 5. Parallel connection and serial connection, 6.
Autonomy, 7. Finitude. Determinism, 8. Order, 9. Non-anticipation, the first
definition, 10. Non-anticipation, the second definition, 11. Time invariance,
12. Symmetry, the first definition, 13. Symmetry, the second definition, 14.
Stability, 15. Fundamental mode, 16. Generator function, Appendix

\bigskip

{\Large 1. Introduction}

\bigskip

We mention three levels of abstraction of digital electrical engineering.

The first level is the descriptive, non-formalized one. The bricks with which
this theory is built are small: logical gates, flip-flops, or bigger:
handshake controls, pipelines, adders, oscillators. The analysis is made
either timeless, with truth tables, or timed (discrete/real) by using
different methods.

The second level was proposed by the author in some previous papers under the
name of delay theory. The fundamental notion is that of delay= the
mathematical model of the delay circuit, consisting in systems of ordinary
and/or differential equations and/or inequalities written on $\mathbf{R}%
\rightarrow\{0,1\}$ functions. For example if the input $u$ and the state $x$
are such functions, the equation%
\[
x(t)=u(t-d)
\]
where $t\in\mathbf{R},\,d\geq0$ is called the ideal delay, while the delays%
\[
\underset{\xi\in\lbrack t-d_{r},t)}{\bigcap}u(\xi)\leq x(t)\leq\underset
{\xi\in\lbrack t-d_{f},t)}{\bigcup}u(\xi)
\]
and respectively%
\[
\overline{x(t-0)}\cdot x(t)\cdot\underset{\xi\in\lbrack t-d_{r},t)}{\bigcap
}u(\xi)\cup x(t-0)\cdot\overline{x(t)}\cdot\underset{\xi\in\lbrack t-d_{f}%
,t)}{\bigcap}\overline{u(\xi)}\cup
\]%
\[
\cup\overline{x(t-0)}\cdot\overline{x(t)}\cdot\overline{\underset{\xi
\in\lbrack t-d_{r},t)}{\bigcap}u(\xi)}\cup x(t-0)\cdot x(t)\cdot
\overline{\underset{\xi\in\lbrack t-d_{f},t)}{\bigcap}\overline{u(\xi)}}=1
\]
are inertial, i.e. non-ideal, where $d_{r}>0,d_{f}>0$. We interpret the last
differential equation (the functions $\overline{x(t-0)}\cdot x(t),x(t-0)\cdot
\overline{x(t)}$ are called the left semi-derivatives of $x\,$) in the next
manner: at each time instant $t$, one of the next conditions is true

\begin{itemize}
\item $x$ was $0$ and now it is $1$ and $u$ was $1$ for sufficiently long
($d_{r}$ tine units)

\item $x$ was $1$ and now it is $0$ and $u$ was $0$ for sufficiently long

\item $x$ was $0$ and now it is $0$ and $u$ was not $1$ for sufficiently long

\item $x$ was $1$ and now it is $1$ and $u$ was not $0$ for sufficiently long
\end{itemize}

With delays and Boolean functions, any asynchronous circuit may be modeled at
the most detailed logical level and this is sometimes an advantage, sometimes
a disadvantage.

The third level of abstraction of digital electrical engineering is the one of
the system theory that is inspired by the delay theory. In fact when the
details that characterize delay theory are a (major) disadvantage, they are
avoided by using asynchronous systems. An asynchronous system $f$ (in the
input-output sense) is a 'black-box', thought as a multivalued function
associating to each input $u:\mathbf{R}\rightarrow\{0,1\}^{m}$ respectively a
set of states $x:\mathbf{R}\rightarrow\{0,1\}^{n},\,x\in f(u)$. The
one-to-many association (in other words: the non-deterministic association)
that $f$ represents is motivated by the fact that the parameters that define
an asynchronous circuit are not known and constant:

\begin{itemize}
\item they are known within the limits given by the precission of the
measurement tools

\item they depend on the temperature and on the power supply, thus they are
variable against time in some ranges of values

\item they depend on the technology that is used, but they differ even if we
compare similar circuits produced in the same technology
\end{itemize}

In the paper we propose to analyze different types of asynchronous systems.

\bigskip

\bigskip{\Large 2. Preliminaries}

We introduce now some notions, notations and preliminary results.

$\mathbf{R}$ is the time set. For $t,d\in\mathbf{R}$, the function $\tau
^{d}:\mathbf{R}\rightarrow\mathbf{R},$ $\tau^{d}(t)=t-d$ is the time
translation with $d$.

We note with $\mathbf{B}$ the set $\{0,1\}$ and let%
\[
P^{\ast}(\mathbf{B}^{m})=\{A|A\subset\mathbf{B}^{m},A\neq\emptyset\}
\]
A signal is a function $w:\mathbf{R}\rightarrow\mathbf{B}$ with the property
that the real unbounded sequence $0\leq t_{0}<t_{1}<t_{2}<...$ exists so that%
\[
w(t)=w(t_{0}-0)\cdot\chi_{(-\infty,t_{0})}(t)\oplus w(t_{0})\cdot\chi_{\lbrack
t_{0},t_{1})}(t)\oplus w(t_{1})\cdot\chi_{\lbrack t_{1},t_{2})}(t)\oplus...
\]
where $\chi_{(\cdot)}$ is the characteristic function. The set of the signals
is noted with $S$ and we note furthermore:%
\[
S^{(m)}=\{u|u:\mathbf{R}\rightarrow\mathbf{B}^{m},u_{i}\in S,i=\overline
{1,m}\},m\geq1
\]%
\[
S^{(0)}=\{0|0:\mathbf{R}\rightarrow\mathbf{B}\}
\]
(the one element set consisting in the null function), respectively%
\[
P^{\ast}(S^{(m)})=\{U|U\subset S^{(m)},U\neq\varnothing\}
\]

For $\lambda\in\mathbf{B}^{m},$ $u\in S^{(m)}$ and $\sigma
:\{1,...,m\}\rightarrow\{1,...,m\}$ bijective, we note%
\[
\overline{\lambda}=(\overline{\lambda_{1}},...,\overline{\lambda_{m}}%
),\quad\overline{u}(t)=(\overline{u_{1}(t)},...,\overline{u_{m}(t)})
\]%
\[
\lambda_{\sigma}=(\lambda_{\sigma(1)},...,\lambda_{\sigma(m)}),\quad
u_{\sigma}(t)=(u_{\sigma(1)}(t),...,u_{\sigma(m)}(t))
\]
\textbf{Lemma} Let $u\in S^{(m)},m\geq1$. The next statements are true

a) If $u$ is not constant we note $t_{0}=\min\{t|u(t-0)\neq u(t)\}$. Then%
\[
\forall d\in\mathbf{R},(u\circ\tau^{d}\in S^{(m)}\Longleftrightarrow
t_{0}+d\geq0)
\]

b) For any $d\geq0$, we have $u\circ\tau^{d}\in S^{(m)}$

c) $(\forall d\in\mathbf{R},$ $u\circ\tau^{d}\in S^{(m)})$
$\Longleftrightarrow$ $u$ is constant.

\begin{proof}
We suppose that the family $u^{0},u^{1},u^{2},...\in\mathbf{B}^{m}$ and the
unbounded sequence $0\leq t_{0}<t_{1}<t_{2}<...$ are chosen so that%
\[
u(t)=u^{0}\cdot\chi_{(-\infty,t_{0})}(t)\oplus u^{1}\cdot\chi_{\lbrack
t_{0},t_{1})}(t)\oplus u^{2}\cdot\chi_{\lbrack t_{1},t_{2})}(t)\oplus...
\]
and if $u$ is not constant, then $t_{0}=\min\{t|u(t-0)\neq u(t)\}$. For any
$d\in\mathbf{R}$ we can write%
\begin{align*}
u\circ\tau^{d}(t)  &  =u(t-d)=\\
&  =u^{0}\cdot\chi_{(-\infty,t_{0})}(t-d)\oplus u^{1}\cdot\chi_{\lbrack
t_{0},t_{1})}(t-d)\oplus u^{2}\cdot\chi_{\lbrack t_{1},t_{2})}(t-d)\oplus...\\
&  =u^{0}\cdot\chi_{(-\infty,t_{0}+d)}(t)\oplus u^{1}\cdot\chi_{\lbrack
t_{0}+d,t_{1}+d)}(t)\oplus u^{2}\cdot\chi_{\lbrack t_{1}+d,t_{2}+d)}%
(t)\oplus...
\end{align*}
The sequence with the general term $t_{k}^{\prime}=t_{k}+d,\,k\in\mathbf{N}$
is unbounded, like $(t_{k})$.

a) Obvious.

b) If $u$ is constant, then $\forall d\in\mathbf{R},$ $u=u\circ\tau^{d}\in
S^{(m)}$ and if $u$ is not constant, then the property results from the fact
that $0\leq d,\,0\leq t_{0}\leq t_{0}+d$ and from a) $\Longleftarrow$.

c) $\Longrightarrow$ We suppose against all reason that $u$ is not constant.
Some $d\in\mathbf{R}$ exists then so that $t_{0}+d<0$ thus, from a) we get
$u\circ\tau^{d}\notin S^{(m)}$, contradiction.
\end{proof}

\bigskip

{\Large 3. Asynchronous systems}

\begin{definition}
The functions $f:S^{(m)}\rightarrow P^{\ast}(S^{(n)})$ are called asynchronous
systems (in the input-output sense), shortly systems. The elements $u\in
S^{(m)}$, respectively $x\in f(u)$ are called inputs, respectively states (or outputs).
\end{definition}

\begin{remark}
The asynchronous systems $f$ are relations of determination between the cause
$u$ and the effect $x\in f(u)$ and our only request is that each cause has
effects: $\forall u,f(u)\neq\varnothing$. When this determination consists in
a system of equations and/or inequalities, $f$ gives for any $u$ the set
$f(u)$ of the solutions of the system (writting systems of equations and/or
inequalities is not the purpose of the present paper, however).

The one-to-many association $u\longmapsto f(u)$ has its origin as we have
already mentioned in the fact that to one cause $u$ there correspond in
general several possible effects $x\in f(u)$ depending on the variations in
ambient temperature, power supply, on the technology etc.
\end{remark}

\begin{example}
In the next examples we have $m=n$ at (1) and $n=1$ at (2),...,(4):%
\begin{equation}
f(u)=\{u\circ\tau^{d}\},d\geq0
\end{equation}%
\begin{equation}
f(u)=\{x|\exists d\geq0,\forall t\geq d,x(t)=u_{i}(t)\},i\in\{1,...,m\}
\end{equation}%
\begin{equation}
f(u)=\{x|\overline{x(t-0)}\cdot x(t)\leq u_{1}(t-d)\cdot...\cdot u_{m}(t-d)\}
\end{equation}%
\begin{equation}
f(u)=\{x|\overline{x(t-0)}\cdot x(t)\leq\underset{\xi\in\lbrack t,t+\delta
_{r}]}{\bigcap}x(\xi),x(t-0)\cdot\overline{x(t)}\leq\underset{\xi\in\lbrack
t,t+\delta_{f}]}{\bigcap}\overline{x(\xi)}\}
\end{equation}
For (1), the fact that $u\in S^{(m)}$ and $d\geq0$ implies $u\circ\tau^{d}\in
S^{(m)}$ has been proved in the Lemma, item b). And at (4) where $\delta
_{r}\geq0,\delta_{f}\geq0$ a system is defined that associates to each input
$u$ the set of all the (inertial) states $x$ having the property that if they
switch from $0$ to $1$ they remain $1$ more than $\delta_{r}$ time units and
if they switch from $1$ to $0$ they remain $0$ more than $\delta_{f}$ time units.
\end{example}

\begin{definition}
Let the set $X\in P^{\ast}(S^{(n)})$ and the systems $f,g:S^{(m)}\rightarrow
P^{\ast}(S^{(n)})$. They define the next systems:

\begin{itemize}
\item $\overline{f}:S^{(m)}\rightarrow P^{\ast}(S^{(n)}),$%
\[
\forall(u_{1},...,u_{m})\in S^{(m)},\overline{f}(u_{1},...,u_{m}%
)=\{\overline{x}|x\in f(u_{1},...,u_{m})\}
\]

\item $f^{(m+1)}:S^{(m+1)}\rightarrow P^{\ast}(S^{(n)}),$%
\[
\forall(u_{1},...,u_{m+1})\in S^{(m+1)},f^{(m+1)}(u_{1},...,u_{m+1}%
)=f(u_{1},...,u_{m})
\]

\item $f_{i\rightarrow j}:S^{(m)}\rightarrow P^{\ast}(S^{(n)})$ is defined for
all $i,j\in\{1,...,m\},i\neq j$ \ by%
\[
\forall(u_{1},...,u_{m})\in S^{(m)},f_{i\rightarrow j}(u_{1},...,\underset
{i}{u_{i}},...,\underset{j}{u_{j}},...,u_{m})=f(u_{1},...,\underset{i}{u_{i}%
},...,\underset{j}{u_{i}},...,u_{m})
\]

\item we suppose that $f$ does not depend on $u_{i},i\in\{1,...,m\}$ i.e. for
all $u\in S^{(m)},f(u_{1},...,u_{i},...,u_{m})=f(u_{1},...,\underset{i}%
{0},...,u_{m})$. Then $f_{\widehat{u}_{i}}:S^{(m-1)}\rightarrow P^{\ast
}(S^{(n)})$ is defined in the next manner:%
\[
\forall(u_{1},...,\widehat{u}_{i},...,u_{m})\in S^{(m-1)},f_{\widehat{u}_{i}%
}(u_{1},...,\widehat{u}_{i},...,u_{m})=f(u_{1},...,\underset{i}{0},...,u_{m})
\]
where $\widehat{u}_{i}$ indicates a missing coordinate

\item if $\forall(u_{1},...,u_{m})\in S^{(m)},f(u_{1},...,u_{m})\cap
X\neq\emptyset$, respectively if $\forall(u_{1},...,u_{m})\in S^{(m)}%
,f(u_{1},...,u_{m})\cap g(u_{1},...,u_{m})\neq\emptyset$, then the systems
$f\cap X,f\cap g:S^{(m)}\rightarrow P^{\ast}(S^{(n)})$ are defined by%
\[
\forall(u_{1},...,u_{m})\in S^{(m)},(f\cap X)(u_{1},...,u_{m})=f(u_{1}%
,...,u_{m})\cap X
\]%
\[
\forall(u_{1},...,u_{m})\in S^{(m)},(f\cap g)(u_{1},...,u_{m})=f(u_{1}%
,...,u_{m})\cap g(u_{1},...,u_{m})
\]

\item $f\cup X,f\cup g:S^{(m)}\rightarrow P^{\ast}(S^{(n)})$,%
\[
\forall(u_{1},...,u_{m})\in S^{(m)},(f\cup X)(u_{1},...,u_{m})=f(u_{1}%
,...,u_{m})\cup X
\]%
\[
\forall(u_{1},...,u_{m})\in S^{(m)},(f\cup g)(u_{1},...,u_{m})=f(u_{1}%
,...,u_{m})\cup g(u_{1},...,u_{m})
\]
\end{itemize}
\end{definition}

\bigskip

{\Large 4. Initial states}

\begin{definition}
\bigskip Let the system $f$. The function $\phi:S^{(m)}\rightarrow P^{\ast
}(\mathbf{B}^{n})$,%
\[
\forall u,\phi(u)=\{x(0-0)|x\in f(u)\}
\]
is called the initial state function of $f$ and the set
\[
\Theta_{f}=\underset{u\in S^{(m)}}{\bigcup}\phi(u)
\]
is called the set of the initial states of $f$.
\end{definition}

\begin{definition}
If $\Theta_{f}=\{x^{0}\}$ i.e. if%
\[
\forall u,\forall x\in f(u),x(0-0)=x^{0}%
\]
then we say that $f$ is initialized and that $x^{0}$ is the initial state of
$f$; otherwise, we say that $f$ is not initialized and that it does not have
an initial state.
\end{definition}

\begin{example}
\bigskip The constant function $S^{(m)}\rightarrow P^{\ast}(S^{(n)})$ equal
with $(x^{0}\}$ is an initialized system whose initial state is $x^{0}$.
\end{example}

\begin{remark}
Many authors prefer to work either with initialized systems, or at least with
constant initial state functions. Our option is for a more general frame
because we want to include in this study the trivial systems $f(u)=\{u\}$ and
other similar systems.
\end{remark}

\begin{theorem}
Let the systems $f,g$ and the set of states $X$. The initial state functions
of the systems $\overline{f},f^{(m+1)},f_{i\rightarrow j},f_{\widehat{u}_{i}%
},f\cap X,f\cap g,f\cup X,f\cup g$ are the next ones:

\begin{itemize}
\item $\overline{\phi}:S^{(m)}\rightarrow P^{\ast}(\mathbf{B}^{n}),$%
\[
\forall(u_{1},...,u_{m})\in S^{(m)},\overline{\phi}(u_{1},...,u_{m}%
)=\{\overline{x^{0}}|x^{0}\in\phi(u_{1},...,u_{m})\}
\]

\item $\phi^{(m+1)}:S^{(m+1)}\rightarrow P^{\ast}(\mathbf{B}^{n}),$%
\[
\forall(u_{1},...,u_{m+1})\in S^{(m+1)},\phi^{(m+1)}(u_{1},...,u_{m+1}%
)=\phi(u_{1},...,u_{m})
\]

\item $\phi_{i\rightarrow j}:S^{(m)}\rightarrow P^{\ast}(\mathbf{B}^{n})$ is
given for all $i,j\in\{1,...,m\},i\neq j$ \ by%
\[
\forall(u_{1},...,u_{m})\in S^{(m)},\phi_{i\rightarrow j}(u_{1},...,\underset
{i}{u_{i}},...,\underset{j}{u_{j}},...,u_{m})=\phi(u_{1},...,\underset
{i}{u_{i}},...,\underset{j}{u_{i}},...,u_{m})
\]

\item if $f$ does not depend on $u_{i},i\in\{1,...,m\}$, then $\phi
_{\widehat{u}_{i}}:S^{(m-1)}\rightarrow P^{\ast}(\mathbf{B}^{n})$ is given by:%
\[
\forall(u_{1},...,\widehat{u}_{i},...,u_{m})\in S^{(m-1)},\phi_{\widehat
{u}_{i}}(u_{1},...,\widehat{u}_{i},...,u_{m})=\phi(u_{1},...,\underset{i}%
{0},...,u_{m})
\]

\item if $\forall(u_{1},...,u_{m})\in S^{(m)},f(u_{1},...,u_{m})\cap
X\neq\emptyset$, respectively if $\forall(u_{1},...,u_{m})\in S^{(m)}%
,f(u_{1},...,u_{m})\cap g(u_{1},...,u_{m})\neq\emptyset$, then $\phi\cap
\Xi,\phi\cap\gamma:S^{(m)}\rightarrow P^{\ast}(\mathbf{B}^{n})$ are:%
\[
\forall(u_{1},...,u_{m})\in S^{(m)},(\phi\cap\Xi)(u_{1},...,u_{m})=\phi
(u_{1},...,u_{m})\cap\Xi
\]%
\[
\forall(u_{1},...,u_{m})\in S^{(m)},(\phi\cap\gamma)(u_{1},...,u_{m}%
)=\phi(u_{1},...,u_{m})\cap\gamma(u_{1},...,u_{m})
\]
where $\forall(u_{1},...,u_{m})\in S^{(m)},\Xi(u_{1},...,u_{m})=\{x(0-0)|x\in
X\}\overset{not}{=}\Xi$

\item $\phi\cup\Xi,\phi\cup\gamma:S^{(m)}\rightarrow P^{\ast}(\mathbf{B}^{n}%
)$,%
\[
\forall(u_{1},...,u_{m})\in S^{(m)},(\phi\cup\Xi)(u_{1},...,u_{m})=\phi
(u_{1},...,u_{m})\cup\Xi
\]%
\[
\forall(u_{1},...,u_{m})\in S^{(m)},(\phi\cup\gamma)(u_{1},...,u_{m}%
)=\phi(u_{1},...,u_{m})\cup\gamma(u_{1},...,u_{m})
\]
\end{itemize}
\end{theorem}

\begin{proof}
These result from the way that the initial state function was introduced at
Definition 5. For example%
\[
\overline{\phi}(u)=\{x(0-0)|x\in\overline{f}(u)\}=\{x(0-0)|\overline{x}\in
f(u)\}=\{\overline{x(0-0)}|x\in f(u)\}=\{\overline{x^{0}}|x^{0}\in\phi(u)\}
\]
\end{proof}

\bigskip

{\Large 5. Parallel connection and serial connection}

\begin{remark}
We shall identify the sets $S^{(m_{1})}\times...\times S^{(m_{p})}$ and
$S^{(m_{1}+...+m_{p})}$ for $m_{1}\geq1,...,m_{p}\geq1$ whose elements are of
the form $(u^{1},...,u^{p})=(u_{1}^{1},...,u_{m_{1}}^{1},...,u_{1}%
^{p},...,u_{m_{p}}^{p}).$ By this identification we ignore the fact that the
argument of $(u^{1},...,u^{p})$ is $(t_{1},...,t_{p})\in\mathbf{R}^{p}$ and
the argument of $(u_{1}^{1},...,u_{m_{1}}^{1},...,u_{1}^{p},...,u_{m_{p}}%
^{p})$ is $t\in\mathbf{R}$ and we just keep in mind the form of the
coordinates of these functions. The convention imposes furthermore the
identification of $P^{\ast}(S^{(n_{1})})\times...\times P^{\ast}(S^{(n_{p})})$
with $P^{\ast}(S^{(n_{1}+...+n_{p})})$. See the Appendix for more details.

These identifications are more meaningful than they might seem at the first
sight because they allow in the next definition that $p$ systems with $p$
different time axes, when connected in parallel, have one time axis.
\end{remark}

\begin{definition}
The parallel connection (or the direct product) of the systems $f^{i}%
:S^{(m_{i})}\rightarrow P^{\ast}(S^{(n_{i})}),i=\overline{1,p}$ is the system
$(f^{1},...,f^{p}):S^{(m_{1}+...+m_{p})}\rightarrow P^{\ast}(S^{(n_{1}%
+...+n_{p})})$ defined by%
\[
\forall(u^{1},...,u^{p})\in S^{(m_{1}+...+m_{p})},(f^{1},...,f^{p}%
)(u^{1},...,u^{p})=(f^{1}(u^{1}),...,f^{p}(u^{p}))
\]
\end{definition}

\begin{definition}
We suppose that $n_{1}+...+n_{p}=m$. The serial connection of the systems
$f,f^{1},...,f^{p}$ is the system $f\circ(f^{1},...,f^{p}):S^{(m_{1}%
+...+m_{p})}\rightarrow P^{\ast}(S^{(n)})$ that is defined by any of the
equivalent statements:%
\[
f\circ(f^{1},...,f^{p})(u^{1},...,u^{p})=\{x|\exists y^{1}\in f^{1}%
(u^{1}),...,\exists y^{p}\in f^{p}(u^{p}),x\in f(y^{1},...,y^{p})\}
\]%
\[
f\circ(f^{1},...,f^{p})(u^{1},...,u^{p})=\underset{(y^{1},...,y^{p})\in
f^{1}(u^{1})\times...\times f^{p}(u^{p})}{\bigcup}f(y^{1},...,y^{p})
\]
\end{definition}

\begin{example}
The system $I:S\rightarrow P^{\ast}(S)$ is defined in the next way%
\begin{equation}
I(u_{i})=\{u_{i}\}
\end{equation}
Then for any $f$ and any $u=(u_{1},...,u_{m})$ we remark that%
\[
(\underset{n}{\underbrace{I,...,I}})\circ f(u)=f\circ(\underset{m}%
{\underbrace{I,...,I}})(u)=f(u)
\]
More general, with the notation $I_{d}:S\rightarrow P^{\ast}(S),d\geq0$%
\begin{equation}
I_{d}(u_{i})=\{u_{i}\circ\tau^{d}\}
\end{equation}
we have%
\begin{equation}
f\circ(\underset{m}{\underbrace{I_{d},...,I_{d}}})(u)=f(u\circ\tau^{d})
\end{equation}%
\begin{equation}
(\underset{n}{\underbrace{I_{d},...,I_{d}}})\circ f(u)=\{x\circ\tau^{d}|x\in
f(u)\}
\end{equation}
Let us consider for example that $f$ represents the set of the solutions of
the system%
\begin{equation}
\overline{x(t-0)}\cdot x(t)\leq\underset{\xi\in\lbrack t-2,\infty)}{\bigcap
}(u_{1}(\xi)\cdot u_{2}(\xi))
\end{equation}%
\begin{equation}
x(t-0)\cdot\overline{x(t)}=0
\end{equation}
that for $(u_{1},u_{2})=(\chi_{\lbrack0,\infty)},\chi_{\lbrack1,\infty)})$ is
given by%
\[
f(u)=\{1\}\cup\{\chi_{\lbrack d^{\prime},\infty)}|d^{\prime}\geq3\}
\]
For $d=1$ in equation $(7)$ we have%
\[
f(u\circ\tau^{1})=\{1\}\cup\{\chi_{\lbrack d^{\prime},\infty)}|d^{\prime}%
\geq4\}
\]
\end{example}

\begin{theorem}
The initial state function of the system $f\circ(f^{1},...,f^{p})$ is the
function $\phi\circ(f^{1},...,f^{p}):S^{(m_{1}+...+m_{p})}\rightarrow P^{\ast
}(\mathbf{B}^{n})$ defined by%
\[
\phi\circ(f^{1},...,f^{p})(u^{1},...,u^{p})=\{x^{0}|\exists y^{1}\in
f^{1}(u^{1}),...,\exists y^{p}\in f^{p}(u^{p}),x^{0}\in\phi(y^{1}%
,...,y^{p})\}
\]
\end{theorem}

\begin{proof}
$\{x(0-0)|x\in f\circ(f^{1},...,f^{p})(u^{1},...,u^{p})\}=\{x(0-0)|\exists
y^{1}\in f^{1}(u^{1}),...,\exists y^{p}\in f^{p}(u^{p}),x\in f(y^{1},...,y^{p})\}$

$=\{x^{0}|\exists y^{1}\in f^{1}(u^{1}),...,\exists y^{p}\in f^{p}%
(u^{p}),x^{0}\in\phi(y^{1},...,y^{p})\}$
\end{proof}

\bigskip

{\Large 6. Autonomy}

\begin{definition}
The system $f$ is autonomous (or free) if it is the constant function%
\[
\exists X,\forall u,f(u)=X
\]
and it is non-autonomous otherwise. The usual notation for the autonomous
system $f$ is $X$.
\end{definition}

\begin{remark}
The autonomous systems are or may be considered to be without input since the
states $x\in X$ are the same for all $u$. Definition 15 is somehow different
from other authors' point of view [1] that consider the autonomous systems be
those systems where the input takes exactly one value and it belongs -in our
formalization- to the one element set $S^{(0)}$. See however Theorem 18.
\end{remark}

\begin{example}
The (absolute inertial) system $f$ that was defined at Example 3 (4) is autonomous.
\end{example}

\begin{theorem}
If $f$ is autonomous, then $\overline{f},f^{(m+1)},f_{i\rightarrow
j},f_{\widehat{u}_{i}}$ are autonomous, $i,j\in\{1,...,m\},i\neq j$ and a
system $g:S^{(0)}\rightarrow P^{\ast}(S^{(n)})$ exists so that $\forall u\in
S^{(m)},\forall u^{\prime}\in S^{(0)},f(u)=g(u^{\prime})$.
\end{theorem}

\begin{proof}
If $f=X$, then $\forall u,\overline{f}(u)=\{\overline{x}|x\in X\}$ etc. We
take $g=f_{\widehat{u}_{1}...\widehat{u}_{m}}.$
\end{proof}

\begin{theorem}
If $\forall u,X\subset f(u)$ then $f\cap X$ is autonomous and if $\forall
u,f(u)\subset X$, then $f\cup X$ is autonomous.
\end{theorem}

\begin{proof}
For all $u,x$ we have%
\[
x\in X\Longleftrightarrow(x\in X\quad and\quad x\in X)\Longrightarrow(x\in
f(u)\quad and\quad x\in X)\Longrightarrow x\in X
\]
in other words $\forall u,X\subset f(u)\cap X\subset X$ and eventually $f\cap
X=X$.
\end{proof}

\begin{theorem}
If $f,g$ are autonomous, then $f\cap g$ and $f\cup g$ are autonoous.
\end{theorem}

\begin{proof}
If $\exists X,\forall u,f(u)=X$ and $\exists Y,\forall u,g(u)=Y$, then
$\forall u,(f\cap g)(u)=X\cap Y$ and $\forall u,(f\cup g)(u)=X\cup Y$.
\end{proof}

\begin{theorem}
The initial state function $\Xi$ of the autonomous system $X$ is constant and
the initial state functions $\overline{\Xi},\,\Xi^{(m+1)},\,\Xi_{i\rightarrow
j},\,\Xi_{\widehat{u}_{i}}$ are also constant, $i,j\in\{1,...,m\},i\neq j.$
\end{theorem}

\begin{proof}
The set $\forall u,\Xi(u)=\{x(0-0)|x\in X\}$ does not depend on $u$.
\end{proof}

\begin{theorem}
Let $f:S^{(m)}\rightarrow P^{\ast}(S^{(n)})$ and $f^{i}:S^{(m_{i})}\rightarrow
P^{\ast}(S^{(n_{i})}),i=\overline{1,p}$, $n_{1}+...+n_{p}=m$ like before. If
$f$ is autonomous, then $f\circ(f^{1},...,f^{p})$ is autonomous. If
$f^{1},...,f^{p}$ are all autonomous, then $f\circ(f^{1},...,f^{p})$ is autonomous.
\end{theorem}

\begin{proof}
If $f=X$, then $f\circ(f^{1},...,f^{p})=X$ and if $f^{1}=X_{1},...,f^{p}%
=X_{p}$, the formula%
\[
\forall(u^{1},...,u^{p})\in S^{(m_{1}+...+m_{p})},f\circ(f^{1},...,f^{p}%
)(u^{1},...,u^{p})=\underset{(y^{1},...,y^{p})\in X_{1}\times...\times X_{p}%
}{\bigcup}f(y^{1},...,y^{p})
\]
proves the desired property.
\end{proof}

\bigskip

{\Large 7. Finitude. Determinism}

\begin{definition}
The system $f$ is finite (deterministic) if it has the property that $\forall
u,f(u)$ has a finite number of elements (a single element); otherwise, it is
called infinite (non-deterministic).
\end{definition}

\begin{remark}
In the situation when $f$ represents the set of the solutions of a system of
equations/inequalities, its determinism coincides with the uniqueness of the solution.

The deterministic systems may be identified with the $S^{(m)}\rightarrow
S^{(n)}$ functions.

Finiteness is useful when, in modeling, we take in consideration the 'worst
case', the 'best case', the 'most frequent' case etc.
\end{remark}

\begin{example}
We have had already several examples of deterministic systems; we just remark
that the Boolean functions $F:\mathbf{B}^{m}\rightarrow\mathbf{B}^{n}$ define
deterministic systems by%
\[
\forall u,f(u)=\{F(u)\}
\]
The direct product $(F^{1},...,F^{p})$ of $F^{i}:\mathbf{B}^{m_{i}}%
\rightarrow\mathbf{B}^{n_{i}},i=\overline{1,p}$ defines the deterministic
system $(f^{1},...,f^{p})$, where%
\[
\forall u^{i}\in S^{(m_{i})},f^{i}(u^{i})=\{F^{i}(u^{i})\},i=\overline{1,p}%
\]
by%
\[
\forall(u^{1},...,u^{p})\in S^{(m_{1}+...+m_{p})},(f^{1},...,f^{p}%
)(u^{1},...,u^{p})=\{(F^{1}(u^{1}),...,F^{p}(u^{p}))\}
\]
\end{example}

\begin{theorem}
If $f$ is finite (deterministic), then $\overline{f},f^{(m+1)}$,
$f_{i\rightarrow j}$ and $f_{\widehat{u}_{i}}$ are finite (deterministic),
where $i,j\in\{1,...,m\},i\neq j$.
\end{theorem}

\begin{proof}
We note with
$\vert$
$\vert$
the number of elements of a finite set and we have $\forall
u,|f(u)|=|\overline{f}(u)|$ etc.
\end{proof}

\begin{theorem}
If one of the systems $f,g$ is finite (deterministic), then $f\cap g$ is
finite (deterministic) and if both are finite, then $f\cup g$ is finite.
\end{theorem}

\begin{proof}
We suppose that $f$ is finite (deterministic) and we infer%
\[
\forall u,|f(u)\cap g(u)|\leq|f(u)|
\]
thus $f\cap g$ is finite (deterministic). If $f,g$ are both finite then we
have%
\[
\forall u,|f(u)\cup g(u)|\leq|f(u)|+|g(u)|
\]
thus $f\cup g$ is finite.
\end{proof}

\begin{theorem}
When $f$ is deterministic, the initial state function $\phi$ fulfills the
property: $\forall u,\phi(u)$ has a single element and the initial state
functions $\overline{\phi},\,\phi^{(m+1)},$ $\phi_{i\rightarrow j},$
$\phi_{\widehat{u}_{i}}$, $i,j\in\{1,...,m\},i\neq j$ are in the same situation.
\end{theorem}

\begin{proof}
The first assertion is obvious and the other statements take into account
Theorem 26.
\end{proof}

\begin{theorem}
If $f,f^{1},...,f^{p}$ are all finite (deterministic), then $f\circ
(f^{1},...,f^{p})$ is finite (deterministic).
\end{theorem}

\begin{proof}
For some arbitrary $(u^{1},...,u^{p})$ we can write%
\[
f\circ(f^{1},...,f^{p})(u^{1},...,u^{p})=\underset{(y^{1},...,y^{p})\in
f^{1}(u^{1})\times...\times f^{p}(u^{p})}{\bigcup}f(y^{1},...,y^{p})
\]
where $f^{1}(u^{1})\times...\times f^{p}(u^{p})$ is finite (has one element).
\end{proof}

\begin{theorem}
$f$ is autonomous and finite (deterministic) if and only if $\exists X\subset
S^{(n)}$ finite (consisting in a single element) so that $\forall u,f(u)=X$.
\end{theorem}

\begin{proof}
Obvious.
\end{proof}

\bigskip

{\Large 8. Order}

\begin{definition}
The next inclusion $f\subset g$ is defined between the systems $f,g$:%
\[
\forall u,f(u)\subset g(u)
\]
\end{definition}

\begin{remark}
$\subset$ is a partial order without first element, but with the last element
represented by the autonomous system $S^{(n)}:S^{(m)}\rightarrow P^{\ast
}(S^{(n)})$,%
\[
\forall u\in S^{(m)},S^{(n)}(u)=S^{(n)}%
\]

The sense of the inclusion $f\subset g$ is that the model offered by $f$ is
more precise, it has more information on the modeled circuit than the model
offered by $g$, in particular the deterministic systems give the maximal
information and the autonomous system $S^{(n)}$ gives the minimal information.
\end{remark}

\begin{example}
We consider the next $S^{(m)}\rightarrow P^{\ast}(S)$ systems%
\[
f_{1}(u)=\{u_{i}\}
\]%
\[
f_{2}(u)=\{x|\forall t\geq0,x(t)=u_{i}(t)\}
\]%
\[
f_{3}(u)=\{x|\exists t^{\prime},\forall t\geq t^{\prime},x(t)=u_{i}(t)\}
\]
where $i\in\{1,...,m\}$. We have $f_{1}\subset f_{2}\subset f_{3}$.
\end{example}

\begin{theorem}
If $f\subset g$, then $\overline{f}\subset\overline{g},\,f^{(m+1)}\subset
g^{(m+1)},\,f_{i\rightarrow j}\subset g_{i\rightarrow j},\,f_{\widehat{u}_{i}%
}\subset g_{\widehat{u}_{i}}$ are true, $i,j\in\{1,...,m\},i\neq j$.
\end{theorem}

\begin{proof}
For example

$\forall(u_{1},...,u_{m}),f_{i\rightarrow j}(u_{1},...,\underset{i}{u_{i}%
},...,\underset{j}{u_{j}},...,u_{m})=f(u_{1},...,\underset{i}{u_{i}%
},...,\underset{j}{u_{i}},...,u_{m})\subset$

$\subset g(u_{1},...,\underset{i}{u_{i}},...,\underset{j}{u_{i}}%
,...,u_{m})=g_{i\rightarrow j}(u_{1},...,\underset{i}{u_{i}},...,\underset
{j}{u_{j}},...,u_{m})$
\end{proof}

\begin{theorem}
For $X\subset S^{(n)}$ and the systems $f,g$, the next inclusions take place:%
\[
f\cap X\subset f\subset f\cup X
\]%
\[
f\cap g\subset f\subset f\cup g
\]
\end{theorem}

\begin{proof}
$\forall u,\forall x,x\in(f\cap g)(u)\Longleftrightarrow x\in f(u)\cap
g(u)\Longleftrightarrow(x\in f(u)$ $and$ $x\in g(u))\Longrightarrow x\in
f(u)\Longrightarrow(x\in f(u)$ $or$ $x\in g(u))\Longleftrightarrow x\in
f(u)\cup g(u)\Longleftrightarrow x\in(f\cup g)(u)$
\end{proof}

\begin{theorem}
If $f\subset g$, then $\forall u,\phi(u)\subset\gamma(u)$.
\end{theorem}

\begin{proof}
For any $u$ we have $\phi(u)=\{x(0-0)|x\in f(u)\}\subset\{x(0-0)|x\in
g(u)\}=\gamma(u)$
\end{proof}

\begin{theorem}
Let the systems $f,g:S^{(m)}\rightarrow P^{\ast}(S^{(n)}),f^{i},g^{i}%
:S^{(m_{i})}\rightarrow P^{\ast}(S^{(n_{i})}),$ $i=\overline{1,p}$ so that
$n_{1}+...+n_{p}=m$. The next implications are true:%
\[
f\subset g\Longrightarrow f\circ(f^{1},...,f^{p})\subset g\circ(f^{1}%
,...,f^{p})
\]%
\[
f^{1}\subset g^{1},...,f^{p}\subset g^{p}\Longrightarrow f\circ(f^{1}%
,...,f^{p})\subset f\circ(g^{1},...,g^{p})
\]
\end{theorem}

\begin{proof}
Let $(u^{1},...,u^{p})$ and $x\in f\circ(f^{1},...,f^{p})(u^{1},...,u^{p})$,
meaning that $y^{1}\in f^{1}(u^{1}),...,y^{p}\in f^{p}(u^{p})$ exist so that
$x\in f(y^{1},...,y^{p})$; because $x\in g(y^{1},...,y^{p})$, we obtain $x\in
g\circ(f^{1},...,f^{p})(u^{1},...,u^{p})$.

On the other hand if we suppose that $x\in f\circ(f^{1},...,f^{p}%
)(u^{1},...,u^{p})$, then $y^{1}\in f^{1}(u^{1}),...,y^{p}\in f^{p}(u^{p})$
exist so that $x\in f(y^{1},...,y^{p})$. We get $y^{1}\in g^{1}(u^{1}%
),...,y^{p}\in g^{p}(u^{p})$ and this implies $x\in f\circ(g^{1}%
,...,g^{p})(u^{1},...,u^{p})$.
\end{proof}

\begin{theorem}
Let the arbitrary sets $X\subset S^{(n)}$,$X_{i}\subset S^{(n_{i}%
)},i=\overline{1,p}$ and the systems $f,g:S^{(m)}\rightarrow P^{\ast}%
(S^{(n)}),\,f^{i},g^{i}:S^{(m_{i})}\rightarrow P^{\ast}(S^{(n_{i}%
)}),i=\overline{1,p}$ so that $n_{1}+...+n_{p}=m$.

\begin{itemize}
\item[a)] If $\forall u,f(u)\cap X\neq\emptyset$, then $\forall(u^{1}%
,...,u^{p}),(f\circ(f^{1},...,f^{p}))(u^{1},...,u^{p})\cap X\neq\emptyset$ and%
\[
(f\cap X)\circ(f^{1},...,f^{p})=(f\circ(f^{1},...,f^{p}))\cap X
\]
If $\forall(u^{1},...,u^{p}),f^{1}(u^{1})\cap X_{1}\neq\emptyset
,...,f^{p}(u^{p})\cap X_{p}\neq\emptyset$, then $\forall(u^{1},...,u^{p}%
),(f\circ(f^{1},...,f^{p}))(u^{1},...,u^{p})\cap(f\circ(X_{1},...,X_{p}%
))\neq\emptyset$ and we can write%
\[
f\circ(f^{1}\cap X_{1},...,f^{p}\cap X_{p})\subset(f\circ(f^{1},...,f^{p}%
))\cap(f\circ(X_{1},...,X_{p}))
\]

\item[b)] If $\forall u,f(u)\cap g(u)\neq\emptyset$, then $\forall
(u^{1},...,u^{p}),(f\circ(f^{1},...,f^{p}))(u^{1},...,u^{p})\cap(g\circ
(f^{1},...,f^{p}))(u^{1},...,u^{p})\neq\emptyset$ and%
\[
(f\cap g)\circ(f^{1},...,f^{p})\subset(f\circ(f^{1},...,f^{p}))\cap
(g\circ(f^{1},...,f^{p}))
\]
If $\forall(u^{1},...,u^{p}),f^{1}(u^{1})\cap g^{1}(u^{1})\neq\emptyset
,...,f^{p}(u^{p})\cap g^{p}(u^{p})\neq\emptyset$, then $\forall(u^{1}%
,...,u^{p}),(f\circ(f^{1},...,f^{p}))(u^{1},...,u^{p})\cap(f\circ
(g^{1},...,g^{p}))(u^{1},...,u^{p})\neq\emptyset$ and%
\[
f\circ(f^{1}\cap g^{1},...,f^{p}\cap g^{p})\subset(f\circ(f^{1},...,f^{p}%
))\cap(f\circ(g^{1},...,g^{p}))
\]

\item[c)] We have%
\[
(f\cup X)\circ(f^{1},...,f^{p})=(f\circ(f^{1},...,f^{p}))\cup X
\]%
\[
f\circ(f^{1}\cup X_{1},...,f^{p}\cup X_{p})\supset(f\circ(f^{1},...,f^{p}%
))\cup(f\circ(X_{1},...,X_{p}))
\]

\item[d)] The next properties are also true:%
\[
(f\cup g)\circ(f^{1},...,f^{p})=(f\circ(f^{1},...,f^{p}))\cup(g\circ
(f^{1},...,f^{p}))
\]%
\[
f\circ(f^{1}\cup g^{1},...,f^{p}\cup g^{p})\supset(f\circ(f^{1},...,f^{p}%
))\cup(f\circ(g^{1},...,g^{p}))
\]
\end{itemize}
\end{theorem}

\begin{proof}
We prove b) and respectively d):

$\forall(u^{1},...,u^{p}),((f\cap g)\circ(f^{1},...,f^{p}))(u^{1}%
,...,u^{p})=\{x|\exists y^{1},...,\exists y^{p},y^{1}\in f^{1}(u^{1})\quad
and$\thinspace$...\,and\quad y^{p}\in f^{p}(u^{p})\quad and\quad x\in
f(y^{1},...,y^{p})\quad and\quad x\in g(y^{1},...,y^{p})\}\subset\{x|\exists
y^{1},...,\exists y^{p},y^{1}\in f^{1}(u^{1})\quad and\,...\,and\quad y^{p}\in
f^{p}(u^{p})\quad and\quad x\in f(y^{1},...,y^{p})\quad and\quad\exists
z^{1},...,\exists z^{p},z^{1}\in f^{1}(u^{1})\quad and...and\quad z^{p}\in
f^{p}(u^{p})\quad and\quad x\in g(z^{1},...,z^{p})\}=((f\circ(f^{1}%
,...,f^{p}))\cap(g\circ(f^{1},...,f^{p})))(u^{1},...,u^{p})$

$\forall(u^{1},...,u^{p}),(f\circ(f^{1}\cap g^{1},...,f^{p}\cap g^{p}%
))(u^{1},...,u^{p})=\{x|\exists y^{1},...,\exists y^{p},y^{1}\in f^{1}%
(u^{1})\quad\,and\quad y^{1}\in g^{1}(u^{1})\quad and...and\quad y^{p}\in
f^{p}(u^{p})\quad and\quad y^{p}\in g^{p}(u^{p})\quad and\quad x\in
f(y^{1},...,y^{p})\}\subset\{x|\exists y^{1},...,\exists y^{p},y^{1}\in
f^{1}(u^{1})\quad and...and\quad y^{p}\in f^{p}(u^{p})\quad and\quad\exists
z^{1},...,\exists z^{p},z^{1}\in g^{1}(u^{1})\quad and...and\quad z^{p}\in
g^{p}(u^{p})\quad and\quad x\in f(y^{1},...,y^{p})\quad and\quad x\in
f(z^{1},...,z^{p})\}=((f\circ(f^{1},...,f^{p}))\cap(f\circ(g^{1}%
,...,g^{p})))(u^{1},...,u^{p})$

respectively

$\forall(u^{1},...,u^{p}),((f\cup g)\circ(f^{1},...,f^{p}))(u^{1}%
,...,u^{p})=\{x|\exists y^{1},...,\exists y^{p},y^{1}\in f^{1}(u^{1})\quad
and\,...\,and\quad y^{p}\in f^{p}(u^{p})\quad and\quad(x\in f(y^{1}%
,...,y^{p})\quad or\quad x\in g(y^{1},...,y^{p}))\}=\{x|\exists y^{1}%
,...,\exists y^{p},y^{1}\in f^{1}(u^{1})\quad and\,...\,and\quad y^{p}\in
f^{p}(u^{p})\quad and\quad x\in f(y^{1},...,y^{p})\quad or\quad y^{1}\in
f^{1}(u^{1})\quad and...and\quad y^{p}\in f^{p}(u^{p})\quad and\quad x\in
g(y^{1},...,y^{p})\}=((f\circ(f^{1},...,f^{p}))\cup(g\circ(f^{1}%
,...,f^{p})))(u^{1},...,u^{p})$

$\forall(u^{1},...,u^{p}),(f\circ(f^{1}\cup g^{1},...,f^{p}\cup g^{p}%
))(u^{1},...,u^{p})=\{x|\exists y^{1},...,\exists y^{p},(y^{1}\in f^{1}%
(u^{1})\quad or\quad y^{1}\in g^{1}(u^{1}))\quad and...and\quad(y^{p}\in
f^{p}(u^{p})\quad or\quad y^{p}\in g^{p}(u^{p}))\quad and\quad x\in
f(y^{1},...,y^{p})\}\supset\{x|\exists y^{1},...,\exists y^{p},y^{1}\in
f^{1}(u^{1})\quad and...and\quad y^{p}\in f^{p}(u^{p})\quad and\quad x\in
f(y^{1},...,y^{p})\quad or\quad y^{1}\in g^{1}(u^{1})\quad and...and\quad
y^{p}\in g^{p}(u^{p})\quad and\quad x\in f(y^{1},...,y^{p})\}=((f\circ
(f^{1},...,f^{p}))\cup(f\circ(g^{1},...,g^{p})))(u^{1},...,u^{p})$
\end{proof}

\begin{remark}
At Theorem 38, the statements from a) and b), respectively the statements from
c) and d) are pairwise similar. To be remarked the asymmetry between the first
statements of a) and b).

On the other hand, for the validity of the next theorem we need that the axiom
of choice holds.
\end{remark}

\begin{theorem}
The next properties of determinism take place:

\begin{itemize}
\item[a)] Any system $g$ includes a deterministic system $f$.

\item[b)] If in the inclusion $f\subset g$ the system $g$ is deterministic,
then $f=g$.
\end{itemize}
\end{theorem}

\begin{proof}
a) For any $u$, the axiom of choice allows choosing from the set $g(u)$ a
point $x$ and defining a selective function $f(u)=\{x\}$. $f$ is a
deterministic system and $\forall u,f(u)\subset g(u).$

b) The formula%
\[
\forall u,f(u)=g(u)
\]
represents the only possibility of choosing $f$ at item a).
\end{proof}

\bigskip

{\Large 9. Non-anticipation, the first definition}

\begin{definition}
$f$ is a non-anticipatory (or causative) system if it satisfies for any $u\in
S^{(m)}$ any $x\in S^{(n)\text{ }}$ and any $d\in\mathbf{R}$ one of the next
equivalent conditions

\begin{itemize}
\item[a)] $x\in f(u)\Longrightarrow(u\circ\tau^{d}\in S^{(m)}\Longrightarrow
x\circ\tau^{d}\in S^{(n)})$

\item[b)] $(x\in f(u)$ $and$ $u\circ\tau^{d}\in S^{(m)})\Longrightarrow
x\circ\tau^{d}\in S^{(n)}$

Otherwise, we say that $f$ is anticipatory, or anti-causative.
\end{itemize}
\end{definition}

\begin{theorem}
The system $f$ is non-anticipatory if and only if $\forall u,\forall x\in
f(u)$ one of the next statements is true:

a) $x$ is constant

b) $x,u$ are both variable and we have%
\[
\min\{t|u(t-0)\neq u(t)\}\leq\min\{t|x(t-0)\neq x(t)\}
\]
thus the first input switch is prior to the first output switch.
\end{theorem}

\begin{proof}
If. When $x$ is constant, $\forall d\in\mathbf{R},x=x\circ\tau^{d}\in S^{(n)}$
and the conclusion of 41 b) is true. And if $x,u$ are not constant, we note%
\begin{align*}
t_{0}  &  =\min\{t|u(t-0)\neq u(t)\}\\
t_{1}  &  =\min\{t|x(t-0)\neq x(t)\}
\end{align*}
In 41 a), $x\in f(u)$ is true, thus $(u\circ\tau^{d}\in S^{(m)}\Longrightarrow
x\circ\tau^{d}\in S^{(n)})$ should be true when $u,x,d$ run in $S^{(m)},f(u)$
and $\mathbf{R}$. The next true statements are equivalent:

$(u\circ\tau^{d}\in S^{(m)}\Longrightarrow x\circ\tau^{d}\in S^{(n)}%
)\overset{Lemma,\,item\,a)}{\Longleftrightarrow}t_{0}+d\geq0\Longrightarrow
t_{1}+d\geq0\Longleftrightarrow t_{0}\leq t_{1}$

Only if. Two possibilities exist of negating the statements

Case I $x$ is variable and $u$ is constant

The hypothesis of 41 b) $(x\in f(u)$ $and$ $u\circ\tau^{d}\in S^{(m)})$ is
true for any $d\in\mathbf{R}$ thus the conclusion is true: $\forall
d\in\mathbf{R},x\circ\tau^{d}\in S^{(n)}$. $x$ is constant from the Lemma,
item c), contradiction

Case II $x$ is variable, $u$ is variable and $t_{0}>t_{1}$

Any $d\in\lbrack-t_{0},-t_{1})$ gives $t_{0}+d\geq0$ and $t_{1}+d<0$, i.e.
from the Lemma item a) we get $u\circ\tau^{d}\in S^{(m)}$ and $x\circ\tau
^{d}\notin S^{(n)}$ contradiction with the non-anticipation of $f$.
\end{proof}

\begin{corollary}
We suppose that $f$ is non-anticipatory and we consider the functions $u,x\in
f(u)$.

a) If $u$ is constant, then $x$ is constant.

b) If $u$ is not constant, then two possibilities exist: either $x$ is
constant, or $x$ is not constant and the next condition%
\[
\min\{t|u(t-0)\neq u(t)\}\leq\min\{t|x(t-0)\neq x(t)\}
\]
is fulfilled,
\end{corollary}

\begin{proof}
a) Special case of Theorem 42, item a), only if.

b) Special case of Theorem 42, item a), only if or coincidence with Theorem 42
item b), only if.
\end{proof}

\begin{example}
We have met non-anticipatory systems at Example 3 (1) and the system $f_{1}$
from Example 33 has the same property. Another case is that of the system $f$
with $\forall u,\forall x\in f(u),x$ is the constant function. The system
defined by the next equation is also non-anticipatory:%
\[
x(t)=\underset{\xi\in(-\infty,t)}{\bigcap}u_{i}(\xi)
\]
where $i\in\{1,...,m\}$ , since for all $u$, either $x$ is constant, or it is
variable with exactly one switch from 1 to 0 and in this case we can write%
\[
\min\{t|x(t-0)\neq x(t)\}=\min\{t|u_{i}(t-0)\neq u_{i}(t)\}\geq\min
\{t|u(t-0)\neq u(t)\}
\]
see Theorem 42, if.
\end{example}

\begin{theorem}
Let $f$ non-anticipatory and $i,j\in\{1,...,m\},i\neq j$. Then $\overline{f},$
$f^{(m+1)},$ $f_{i\rightarrow j},$ $f_{\widehat{u}_{i}}$ are non-anticipatory.
\end{theorem}

\begin{proof}
Let $u,x\in\overline{f}(u)$ and $d\in\mathbf{R}$ arbitrary so that $u\circ
\tau^{d}\in S^{(m)}$. From the definition of $\overline{f}$ we have that
$\overline{x}\in f(u)$ and because $f$ is non-anticipatory $\overline{x}%
\circ\tau^{d}\in S^{(n)}$ holds and this is equivalent with any of%
\[
\min\{t|\overline{x}(t-d-0)\neq\overline{x}(t-d)\}\geq0
\]%
\[
\min\{t|x(t-d-0)\neq x(t-d)\}\geq0
\]%
\[
x\circ\tau^{d}\in S^{(n)}%
\]
$\overline{f}$ is non-anticipatory.

The fact that%
\[
(x\in f_{\widehat{u}_{i}}(u)\quad and\quad u\circ\tau^{d}\in S^{(m)}%
)\Longrightarrow(x\in f(u)\quad and\quad u\circ\tau^{d}\in S^{(m)}%
)\Longrightarrow x\circ\tau^{d}\in S^{(n)}%
\]
proves that $f_{\widehat{u}_{i}}$ is non-anticipatory.
\end{proof}

\begin{theorem}
The next statements are equivalent for the system $f$:

\begin{itemize}
\item[a)] $f$ is autonomous and non-anticipatory

\item[b)] $\exists X,\forall u,f(u)=X$ and $\forall x\in X,x$ is the constant function
\end{itemize}
\end{theorem}

\begin{proof}
$a)\Longrightarrow b)\quad$If $\exists X,\forall u,f(u)=X$ we suppose against
all reason that $\exists x\in X$ which is not constant and let $t_{1}\geq0$
with $x(t_{1}-0)\neq x(t_{1})$. The existence of an $u$ so that for some
$t_{0}>t_{1}$ we should have $\forall t<t_{0},u(t)=u(0-0)$ and $u(t_{0}-0)\neq
u(t_{0})$ together with the hypothesis of non-anticipation of $f$ give a
contradiction, see Theorem 42 b), only if.

$b)\Longrightarrow a)\quad$The property is true because if $x\in X$ is
constant, then $\forall d\in\mathbf{R},x\circ\tau^{d}\in S^{(n)}$.
\end{proof}

\begin{theorem}
Let the systems $f,g$ and $X\subset S^{(n)}$. If $f$ is non-anticipatory, then

\begin{itemize}
\item[a)] $f\cap X$ and $f\cap g$ are non-anticipatory

\item[b)] $f\cup X$ is non-anticipatory if and only if $X$ understood as
autonomous system is non-anticipatory and $f\cup g$ is non-anticipatory if and
only if $g$ is non-anticipatory.
\end{itemize}
\end{theorem}

\begin{proof}
The implication $\forall u,\forall x,\forall d\in\mathbf{R}$
\[
x\in f(u)\cap g(u)\Longrightarrow x\in f(u)\Longrightarrow(u\circ\tau^{d}\in
S^{(m)}\Longrightarrow x\circ\tau^{d}\in S^{(n)})
\]
shows the validity of a). At b), the supposition that $f,f\cup g$ are
non-anticipatory and $g$ is anticipatory gives%
\[
\exists u,\exists x\in g(u)-f(u),\exists d\in\mathbf{R},u\circ\tau^{d}\in
S^{(m)}\quad and\quad x\circ\tau^{d}\notin S^{(n)}%
\]
contradiction.
\end{proof}

\begin{theorem}
If $f,f^{1},...,f^{p}$ defined like previously are non-anticipatory, then
$f\circ(f^{1},...,f^{p})$ is non-anticipatory.
\end{theorem}

\begin{proof}
We suppose that $x\in f\circ(f^{1},...,f^{p})(u^{1},...,u^{p})$ and
$(u^{1}\circ\tau^{d},...,u^{p}\circ\tau^{d})\in S^{(m_{1}+...+m_{p})}$
resulting the existence of $y^{1}\in f^{1}(u^{1}),...,y^{p}\in f^{p}(u^{p})$
so that $x\in f(y^{1},...,y^{p})$. Because $f^{1},...,f^{p}$ are
non-anticipatory, we get $y^{1}\circ\tau^{d}\in S^{(n_{1})},...,y^{p}\circ
\tau^{d}\in S^{(n_{p})}$ and from the fact that $f$ is non-anticipatory, we
have $x\circ\tau^{d}\in S^{(n)}$ so that $f\circ(f^{1},...,f^{p})$ has
resulted to be non-anticipatory.
\end{proof}

\begin{theorem}
If $f$ is a non-anticipatory system, then any system $g\subset f$ is non-anticipatory.
\end{theorem}

\begin{proof}
$(x\in g(u)\quad and\quad u\circ\tau^{d}\in S^{(m)})\Longrightarrow(x\in
f(u)\quad and\quad u\circ\tau^{d}\in S^{(m)})\Longrightarrow x\circ\tau^{d}\in
S^{(n)}$
\end{proof}

\bigskip

{\Large 10. Non-anticipation, the second definition}

\begin{definition}
The system $f$ is non-anticipatory, or causative if%
\[
\forall t_{1},\forall u,\forall v,(u|_{(-\infty,t_{1})}=v|_{(-\infty,t_{1}%
)})\Longrightarrow(\forall x\in f(u),\exists y\in f(v),x|_{(-\infty,t_{1}%
)}=y|_{(-\infty,t_{1})})
\]
and anticipatory, or anti-causative otherwise.
\end{definition}

\begin{remark}
This is another perspective on non-anticipation than the previous one and the
two notions are independent logically. The definition states that for any
$t_{1}$ any $u$ and any $x\in f(u)$, the restriction $x|_{(-\infty,t_{1})}$
depends only on the restriction $u|_{(-\infty,t_{1})}$ and is independent on
the values of $u(t),t\geq t_{1}$.

A variant of Definition 50 exists, resulted by the replacement of the interval
$(-\infty,t_{1})$ with $(-\infty,t_{1}]$.
\end{remark}

\begin{example}
Let's consider the next systems%
\[
f(u)=\{\chi_{\lbrack0,1)}\oplus u_{1}\cdot\chi_{\lbrack2,\infty)}\}
\]%
\[
g(u)=%
\genfrac{\{}{.}{0pt}{}{\{1\},\text{ }if\text{ }u_{1}=\chi_{\lbrack0,\infty
)}}{\{u_{1}\},\text{ }otherwise}%
\]
$f(u)$ is non-anticipatory in the sense of Definition 50, but it is
anticipatory in the sense of Definition 41 because for $u_{1}(t)=\chi
_{\lbrack2,\infty)}(t)$ the contradiction $u_{1}\circ\tau^{-2}=\chi
_{\lbrack0,\infty)}\in S,x\circ\tau^{-2}=\chi_{\lbrack-2,-1)}\oplus
\chi_{\lbrack0,\infty)}\notin S$ is obtained. $g(u)$ is anticipatory in the
sense of Definition 50, because for $t_{1}=1,u_{1}=\chi_{\lbrack0,\infty
)},v_{1}=\chi_{\lbrack0,2)}$ the contradiction $1|_{(-\infty,1)}\neq
\chi_{\lbrack0,2)}|_{(-\infty,1)}$ is obtained; it is non-anticipatory in the
sense of Definition 41 however.
\end{example}

\begin{theorem}
Let $f$ a non-anticipatory system (Definition 50). Then $\overline{f},$
$f^{(m+1)},$ $f_{i\rightarrow j},$ $f_{\widehat{u}_{i}}$ are non-anticipatory,
with $i,j\in\{1,...,m\},i\neq j$.
\end{theorem}

\begin{proof}
Let $t_{1},u,v$ and $\overline{x}\in f(u)$ arbitrary so that $u|_{(-\infty
,t_{1})}=v|_{(-\infty,t_{1})}$; the hypothesis that $f$ is non-anticipatory
gives the existence of $\overline{y}\in f(v)$ so that $\overline{x}%
|_{(-\infty,t_{1})}=\overline{y}|_{(-\infty,t_{1})}$ i.e. $x|_{(-\infty
,t_{1})}=y|_{(-\infty,t_{1})}$. These show that $\overline{f}$ is non-anticipatory.

We consider $t_{1},u_{1},...,u_{m+1},v_{1},...,v_{m+1}$ and $x\in
f^{(m+1)}(u_{1},...,u_{m+1})=f(u_{1},...,u_{m})$ arbitrary, so that
$(u_{1},...,u_{m+1})|_{(-\infty,t_{1})}=(v_{1},...,v_{m+1})|_{(-\infty,t_{1}%
)}$. From the fact that $f$ is non-anticipatory we have the existence of $y\in
f(v_{1},...,v_{m})=f^{(m+1)}(v_{1},...,v_{m+1})$ so that $x|_{(-\infty,t_{1}%
)}=y|_{(-\infty,t_{1})}$ i.e. $f^{(m+1)}$ is non-anticipatory.

The part of the proof corresponding to $f_{i\rightarrow j}$ and $f_{\widehat
{u}_{i}}$ is similar.
\end{proof}

\begin{theorem}
If $f,g$ are non-anticipatory systems, then $f\cup g$ is non-anticipatory.
\end{theorem}

\begin{proof}
Let $t_{1},u,v$ and $x\in f(u)\cup g(u)$ arbitrary so that $u|_{(-\infty
,t_{1})}=v|_{(-\infty,t_{1})}$. If for example $x\in f(u)$, then the fact that
$f$ is non-anticipatory shows the existence of $y\in f(v)$ so that
$x|_{(-\infty,t_{1})}=y|_{(-\infty,t_{1})}$; we conclude that $y\in f(u)\cup
g(u)$ exists with $x|_{(-\infty,t_{1})}=y|_{(-\infty,t_{1})}$.
\end{proof}

\begin{theorem}
If $f$ is non-anticipatory, then its initial state function $\phi$ satisfies%
\[
\forall u,\forall v,u(0-0)=v(0-0)\Longrightarrow\phi(u)=\phi(v)
\]
\end{theorem}

\begin{proof}
Let $u,v$ arbitrary so that $u(0-0)=v(0-0)$, thus some $t_{1}$ exists with
$u|_{(-\infty,t_{1})}=v|_{(-\infty,t_{1})}$. From the non-anticipation of $f$
we get $\forall x\in f(u),\exists y\in f(v),x|_{(-\infty,t_{1})}%
=y|_{(-\infty,t_{1})}$ thus $\forall x^{0}\in\phi(u)$ we have that $x^{0}%
\in\phi(v)$.
\end{proof}

\begin{theorem}
If $f,f^{1},...,f^{p}$ are non-anticipatory systems, then $f\circ
(f^{1},...,f^{p})$ is non-anticipatory.
\end{theorem}

\begin{proof}
Let $u^{1},...,u^{p},v^{1},...,v^{p}$ and $t_{1}$ arbitrary with
\[
u^{1}|_{(-\infty,t_{1})}=v^{1}|_{(-\infty,t_{1})},...,u^{p}|_{(-\infty,t_{1}%
)}=v^{p}|_{(-\infty,t_{1})}%
\]
and $x\in(f\circ(f^{1},...,f^{p}))(u^{1},...,u^{p})$ arbitrary also, thus
$y^{1}\in f^{1}(u^{1}),...,y^{p}\in f^{p}(u^{p})$ exist so that $x\in
f(y^{1},...,y^{p})$. Because $f^{1},...,f^{p}$ are non-anticipatory, $z^{1}\in
f^{1}(v^{1}),...,z^{p}\in f^{p}(v^{p})$ exist so that
\[
y^{1}|_{(-\infty,t_{1})}=z^{1}|_{(-\infty,t_{1})},...,y^{p}|_{(-\infty,t_{1}%
)}=z^{p}|_{(-\infty,t_{1})}%
\]
and because $f$ is non-anticipatory we get the existence of $x^{\prime}\in
f(z^{1},...,z^{p})$ with $x|_{(-\infty,t_{1})}=x^{\prime}|_{(-\infty,t_{1})}$.
$f\circ(f^{1},...,f^{p})$ is non-anticipatory.
\end{proof}

\begin{theorem}
Any autonomous system $X\subset S^{(n)}$ is non-anticipatory.
\end{theorem}

\begin{proof}
For any $t_{1}$ we have
\[
\forall x\in X,\exists y\in X,x|_{(-\infty,t_{1})}=y|_{(-\infty,t_{1})}%
\]
thus the conclusion of Definition 50 is true.
\end{proof}

\begin{corollary}
If $f$ is non-anticipatory and $X\subset S^{(n)}$, the system $f\cup X$ is non-anticipatory.
\end{corollary}

\begin{proof}
The result follows from Theorem 54 and Theorem 57.
\end{proof}

\begin{theorem}
If $f$ is a deterministic system (understood as $S^{(m)}\rightarrow S^{(n)}$
function), then the next statements are equivalent:

\begin{itemize}
\item[a)] $f$ is non-anticipatory

\item[b)] $\forall t_{1},\forall u,\forall v,(u|_{(-\infty,t_{1}%
)}=v|_{(-\infty,t_{1})})\Longrightarrow(f(u)|_{(-\infty,t_{1})}%
=f(v)|_{(-\infty,t_{1})})$
\end{itemize}
\end{theorem}

\begin{proof}
Obvious.
\end{proof}

\begin{remark}
Proving that $f,g$ non-anticipatory implies that $f\cap g$ is non-anticipatory
was unsuccesful. This leaves open the problem of finding two non-anticipatory
systems $f,g$ so that $\forall u,f(u)\cap g(u)\neq\emptyset$ and $f\cap g$ is anticipatory.
\end{remark}

\bigskip

{\Large 11. Time invariance}

\begin{definition}
The system $f$ is time invariant if $\forall u\in S^{(m)},\forall x\in
S^{(n)},\forall d\in\mathbf{R},$ one of the next equivalent statements is fulfilled:

\begin{itemize}
\item[a)] $(u\circ\tau^{d}\in S^{(m)}$ $and$ $x\in f(u))\Longrightarrow
(x\circ\tau^{d}\in S^{(n)}$ $and$ $x\circ\tau^{d}\in f(u\circ\tau^{d})) $

\item[b)] $((u\circ\tau^{d}\in S^{(m)}$ $and$ $x\in f(u))$ $\Longrightarrow
x\circ\tau^{d}\in S^{(n)})$ $and$ $((u\circ\tau^{d}\in S^{(m)}$ $and$ $x\in
f(u))$ $\Longrightarrow x\circ\tau^{d}\in f(u\circ\tau^{d}))$

If the previous property is not true, then $f$ is called time variable.
\end{itemize}
\end{definition}

\begin{remark}
If the signals would have been defined by replacing the request of existence
of an initial time instant $t_{0}\geq0$ with the existence of an arbitrary
initial time instant $t_{0}$, then time invariance would have simply been
defined by $\forall u,\forall x,\forall d,(x\in f(u)\Longrightarrow x\circ
\tau^{d}\in f(u\circ\tau^{d}))$. The way that $S$ was defined however, it is
tightly related with the first definition of non-anticipation: time invariance
is the property of the non-anticipatory systems (Definition 41) of satisfying
$x\circ\tau^{d}\in f(u\circ\tau^{d})$ whenever $u\circ\tau^{d}\in S^{(m)}$ and
$x\in f(u)$ hold.
\end{remark}

\begin{example}
We analize two deterministic systems.

a) We show that $f(u)=\{u_{i}\circ\tau^{d^{\prime}}\}$ is time invariant,
where $i\in\{1,...,m\}$ and $d^{\prime}\geq0$. The hypothesis $u\circ\tau
^{d}\in S^{(m)}$ states that $u_{i}\circ\tau^{d}\in S$, from where
$(u_{i}\circ\tau^{d})\circ\tau^{d^{\prime}}\in S$ and we have

- $(u_{i}\circ\tau^{d})\circ\tau^{d^{\prime}}=u_{i}\circ\tau^{d+d^{\prime}%
}=(u_{i}\circ\tau^{d^{\prime}})\circ\tau^{d}\in S$
\ \ $\ \ \ \ \ \ \ \ (x\circ\tau^{d}\in S)$

- $(u_{i}\circ\tau^{d^{\prime}})\circ\tau^{d}=(u_{i}\circ\tau^{d})\circ
\tau^{d^{\prime}}$
\ \ \ \ \ \ \ \ \ \ \ \ \ \ \ \ \ \ \ \ \ \ \ \ \ \ \ \ \ \ \ \ \ \ \ \ \ $(x\circ
\tau^{d}\in f(u\circ\tau^{d}))$

b) Let the system defined by the equation%
\begin{equation}
\,x(t)=\underset{\xi\rightarrow\infty}{\lim}\underset{\omega\in(\xi,\infty
)}{\bigcup}(u_{1}(\omega)\cdot...\cdot u_{m}(\omega))
\end{equation}
(the function in $\xi:\,\underset{\omega\in(\xi,\infty)}{\bigcup}(u_{1}%
(\omega)\cdot...\cdot u_{m}(\omega))$ switches at most once from 1 to 0 for
all $u$, thus the limit $\underset{\xi\rightarrow\infty}{\lim}\underset
{\omega\in(\xi,\infty)}{\bigcup}(u_{1}(\omega)\cdot...\cdot u_{m}(\omega))$
always exists and (11) defines a system indeed). Because $x$ is the constant
function, $x\circ\tau^{d}\in S$ is true for any $d$, thus the system is
non-anticipatory in the sense of Definition 41. By observing that for any
$d\in\mathbf{R},$%
\[
\underset{\xi\rightarrow\infty}{\lim}\underset{\omega\in(\xi,\infty)}{\bigcup
}(u_{1}(\omega-d)\cdot...\cdot u_{m}(\omega-d))=\text{ }\underset
{\xi\rightarrow\infty}{\lim}\underset{\omega\in(\xi,\infty)}{\bigcup}%
(u_{1}(\omega)\cdot...\cdot u_{m}(\omega))=x(t)=x(t-d)
\]
the second statement from Definition 61 b) results. The system is time invariant.
\end{example}

\begin{theorem}
Let $f$ time invariant. The next equivalence holds:%
\[
\forall u,\forall x,\forall d\geq0,x\in f(u)\Longleftrightarrow x\circ\tau
^{d}\in f(u\circ\tau^{d})
\]
\end{theorem}

\begin{proof}
$\Longrightarrow$ The statements $u\circ\tau^{d}\in S^{(m)}$ and $x\in f(u)$
are both true. We apply the time invariance of $f$.

$\Longleftarrow$ $(u\circ\tau^{d})\circ\tau^{-d}\in S^{(m)}$ and $x\circ
\tau^{d}\in f(u\circ\tau^{d})$ are true. We apply the time invariance of $f$
again and we get $(x\circ\tau^{d})\circ\tau^{-d}\in f((u\circ\tau^{d}%
)\circ\tau^{-d})$.
\end{proof}

\begin{theorem}
Let $f$ time invariant and $i,j\in\{1,...,m\},i\neq j.$ $\overline{f},$
$f^{(m+1)},$ $f_{i\rightarrow j},$ $f_{\widehat{u}_{i}}$ are time invariant.
\end{theorem}

\begin{proof}
$\overline{f},f^{(m+1)},f_{i\rightarrow j},f_{\widehat{u}_{i}}$ are
non-anticipatory (Definition 41) from Theorem 45. From the truth of the
implication%
\[
(u\circ\tau^{d}\in S^{(m)}\text{ }and\text{ }x\in f(u))\Longrightarrow
x\circ\tau^{d}\in f(u\circ\tau^{d})
\]
for all $u,x$ and $d$ we get the truth of%
\[
(u\circ\tau^{d}\in S^{(m)}\text{ }and\text{ }\overline{x}\in
f(u))\Longrightarrow\overline{x}\circ\tau^{d}\in f(u\circ\tau^{d})
\]
thus $\overline{f}$ is time invariant.

This part of the proof brings nothing new in the other three cases.
\end{proof}

\begin{theorem}
If $f,g$ are time invariant, then $f\cap g,f\cup g$ are time invariant.
\end{theorem}

\begin{proof}
$f\cap g,f\cup g$ are non-anticipatory (Definition 41) from Theorem 47. From
the truth for all $u,x,d$ of%
\[
(u\circ\tau^{d}\in S^{(m)}\text{ }and\text{ }x\in f(u))\Longrightarrow
x\circ\tau^{d}\in f(u\circ\tau^{d})
\]%
\[
(u\circ\tau^{d}\in S^{(m)}\text{ }and\text{ }x\in g(u))\Longrightarrow
x\circ\tau^{d}\in g(u\circ\tau^{d})
\]
we infer with simple computations that%
\[
(u\circ\tau^{d}\in S^{(m)}\text{ }and\text{ }x\in(f\cap g)(u))\Longrightarrow
x\circ\tau^{d}\in(f\cap g)(u\circ\tau^{d})
\]%
\[
(u\circ\tau^{d}\in S^{(m)}\text{ }and\text{ }x\in(f\cup g)(u))\Longrightarrow
x\circ\tau^{d}\in(f\cup g)(u\circ\tau^{d})
\]
are fulfilled.
\end{proof}

\begin{theorem}
We suppose that $f,f^{1},...,f^{p}$ are time invariant. Then $f\circ
(f^{1},...,f^{p})$ is time invariant.
\end{theorem}

\begin{proof}
$f\circ(f^{1},...,f^{p})$ is non-anticipatory (Definition 41), as resulting
from Theorem 48. Let now $(u^{1},...,u^{p}),$ $x\in f\circ(f^{1}%
,...,f^{p})(u^{1},...,u^{p})$ arbitrary and $\ y^{1}\in f^{1}(u^{1}%
),...,y^{p}\in f^{p}(u^{p})$ \ so that $x\in f(y^{1},...,y^{p})$. The
hypothesis states that%
\[
u^{1}\circ\tau^{d}\in S^{(m_{1})},...,u^{p}\circ\tau^{d}\in S^{(m_{p})}%
\]
are true and from the time invariance of $f^{1},...,f^{p}$ we get that%
\[
y^{1}\circ\tau^{d}\in f^{1}(u^{1}\circ\tau^{d}),...,y^{p}\circ\tau^{d}\in
f^{p}(u^{p}\circ\tau^{d})
\]
are true. But $f$ is time invariant itself thus $x\circ\tau^{d}\in
f(y^{1}\circ\tau^{d},...,y^{p}\circ\tau^{d})$. $f\circ(f^{1},...,f^{p})$ is
time invariant.
\end{proof}

\begin{theorem}
The next statements are equivalent:

\begin{itemize}
\item[a)] $f$ is autonomous and time invariant

\item[b)] $\exists X,f=X$ and $\forall x\in X,x$ is the constant function.
\end{itemize}
\end{theorem}

\begin{proof}
$a)\Longrightarrow b)$ $f$ is autonomous and non-anticipatory (Definition 41)
thus $b)$ is true from Theorem 46$.$

$b)\Longrightarrow a)$ $f$ is autonomous and non-anticipatory from Theorem 46.
Furthermore the truth of%
\[
(u\circ\tau^{d}\in S^{(m)}\quad and\quad x\in X)\Longrightarrow x\circ\tau
^{d}\in X
\]
(because $x=x\circ\tau^{d}$ when $x$ is constant) shows the validity of $a)$.
\end{proof}

\begin{corollary}
If $f$ is time invariant and $X$ satisfies $\forall x\in X,x$ is the constant
function, then $f\cap X,$ $f\cup X$ are time invariant.
\end{corollary}

\begin{proof}
This results from Theorem 66 and Theorem 68.
\end{proof}

\bigskip

{\Large 12. Symmetry, the first definition}

\begin{definition}
The Boolean function $F:\mathbf{B}^{m}\rightarrow\mathbf{B}^{n}$ is called
(coordinatewise) symmetrical if for any bijection $\sigma
:\{1,...,m\}\rightarrow\{1,...,m\}$ we have%
\[
\forall\lambda\in\mathbf{B}^{m},F(\lambda)=F(\lambda_{\sigma})
\]
and asymmetrical otherwise.
\end{definition}

\begin{definition}
The system $f$ is (coordinatewise) symmetrical if for any bijection $\sigma$
we have%
\[
\forall u\in S^{(m)},f(u)=f(u_{\sigma})
\]
and it is asymmetrical otherwise.
\end{definition}

\begin{example}
All the systems with $m=1$ are trivially symmetrical and the systems from
Example 3 (3), (4), respectively from Example 63 b) are also symmetrical. If
$F:\mathbf{B}^{m}\rightarrow\mathbf{B}^{n}$ is a symmetrical function, then
the deterministic system induced by $F$ (Example 25) is symmetrical. The
system%
\[
f(u)=\{x|x(t)\geq u_{1}(t)\cdot...\cdot u_{m}(t)\}
\]
is symmetrical too.
\end{example}

\begin{theorem}
$f$ is symmetrical implies that $\overline{f}$ is symmetrical.
\end{theorem}

\begin{proof}
$\overline{f}(u)=\{\overline{x}|x\in f(u)\}=\{\overline{x}|x\in f(u_{\sigma
})\}=\overline{f}(u_{\sigma})$ are true for all $\sigma$ and $u$.
\end{proof}

\begin{theorem}
Let $f,g$ symmetrical systems. Then $f\cap g,f\cup g$ are symmetrical systems.
\end{theorem}

\begin{proof}
We can write for $\sigma,u$ and $x$ arbitrary:

$x\in(f\cap g)(u)\Longleftrightarrow x\in f(u)\quad and\quad x\in
g(u)\Longleftrightarrow x\in f(u_{\sigma})\quad and\quad x\in g(u_{\sigma
})\Longleftrightarrow x\in(f\cap g)(u_{\sigma})$

The proof for the reunion is similar.
\end{proof}

\begin{theorem}
If $f$ is symmetrical, then $\phi$ is symmetrical.
\end{theorem}

\begin{proof}
For any $\sigma$ and $u$ we have $\phi(u)=\{x(0-0)|x\in f(u)\}=\{x(0-0)|x\in
f(u_{\sigma})\}=\phi(u_{\sigma})$.
\end{proof}

\begin{remark}
If $f^{1},...,f^{p}$ are symmetrical systems, then the next symmetry relation
holds%
\[
f\circ(f^{1},...,f^{p})(u^{1},...,u^{p})=f\circ(f^{\sigma^{\prime}%
(1)},...,f^{\sigma^{\prime}(p)})(u_{\sigma_{\sigma^{\prime}(1)}}%
^{\sigma^{\prime}(1)},...,u_{\sigma_{\sigma^{\prime}(p)}}^{\sigma^{\prime}%
(p)})
\]
where $\sigma_{i}:\{1,...,m_{i}\}\rightarrow\{1,...,m_{i}\},i=\overline{1,p}$
and $\sigma^{\prime}:\{1,...,p\}\rightarrow\{1,...,p\}$ are bijections. We
observe that $f\circ(f^{1},...,f^{p})$ is not a symmetrical system in general.
\end{remark}

\begin{theorem}
If $f$ is autonomous, then it is symmetrical.
\end{theorem}

\begin{proof}
$\exists X,\forall u,f(u)=X$ implies for any bijection $\sigma
:\{1,...,m\}\rightarrow\{1,...,m\}$ that $f(u_{\sigma})=X$
\end{proof}

\begin{corollary}
If $f$ is symmetrical, then $f\cap X,f\cup X$ are symmetrical.
\end{corollary}

\begin{proof}
This fact results from Theorem 74 and Theorem 77.
\end{proof}

\bigskip

{\Large 13. Symmetry, the second definition}

\begin{definition}
The function $F:\mathbf{B}^{m}\rightarrow\mathbf{B}^{n}$ is called symmetrical
(in the rising-falling sense) if%
\[
\forall\lambda\in\mathbf{B}^{m},F(\lambda)=\overline{F(\overline{\lambda})}%
\]
and asymmetrical otherwise.
\end{definition}

\begin{definition}
The system $f$ is symmetrical (in the rising-falling sense) if
\[
\forall u,f(u)=\overline{f}(\overline{u})
\]
and respectively asymmetrical otherwise.
\end{definition}

\begin{remark}
This type of symmetry of $f$ states that the form of $x$ under the input $u$
coincides with the form of $\overline{x}$ under the input $\overline{u}$ and
the terminology of rising-falling symmetry is due to the fact that while
$x(t)$ switches at the time instant $t$ in the rising (falling) sense,
$\overline{x}(t)$ switches at the time instant $t$ in the falling (rising)
sense:%
\[
\forall i\in\{1,...,n\},\overline{x_{i}(t-0)}\cdot x_{i}(t)=\overline
{x_{i}(t-0)}\cdot\overline{\overline{x_{i}(t)}},\text{ }x_{i}(t-0)\cdot
\overline{x_{i}(t)}=\overline{\overline{x_{i}(t-0)}}\cdot\overline{x_{i}(t)}%
\]
\end{remark}

\begin{example}
Some examples of symmetrical functions $F(\lambda)$ (Definition 79) are the
affine functions: $\lambda_{i},i=\overline{1,m}$, $\lambda_{i_{1}}%
\oplus\lambda_{i_{2}}\oplus\lambda_{i_{3}},$ $i_{1},i_{2},i_{3}\in
\{1,...,m\}$, etc. The symmetrical Boolean functions define symmetrical
deterministic systems, for example $F:\mathbf{B}^{3}\rightarrow\mathbf{B}%
,F(\lambda_{1},\lambda_{2},\lambda_{3})=\lambda_{1}\oplus\lambda_{2}%
\oplus\lambda_{3}$ is symmetrical and it defines the symmetrical deterministic
system $f(u)=\{u_{1}\oplus u_{2}\oplus u_{3}\}$.

Let now the non-deterministic system $f(u)=\{u_{1}\cdot u_{2}\}\cup\{u_{1}\vee
u_{2}\}$. The satisfaction of the Morgan laws%
\[
x(t)=u_{1}(t)\cdot u_{2}(t)\Longleftrightarrow\overline{x(t)}=\overline
{u_{1}(t)}\vee\overline{u_{2}(t)}%
\]%
\[
x(t)=u_{1}(t)\vee u_{2}(t)\Longleftrightarrow\overline{x(t)}=\overline
{u_{1}(t)}\cdot\overline{u_{2}(t)}%
\]
shows that it is symmetrical.
\end{example}

\begin{theorem}
If $f$ is symmetrical, then $\overline{f},f^{(m+1)},f_{i\rightarrow j}$ and
$f_{\widehat{u}_{i}}$ are symmetrical for all $i,j\in\{1,...,m\},i\neq j$.
\end{theorem}

\begin{proof}
The conditions of symmetry%
\[
\forall u,f(u)=\overline{f}(\overline{u})
\]%
\[
\forall u,\overline{f}(u)=f(\overline{u})
\]
of $f$ and $\overline{f}$ are equivalent, proving the first statement of the theorem.

We suppose that $f$ is symmetrical. For any $(u_{1},...,u_{m+1})$ we can write%
\[
f^{(m+1)}(u_{1},...,u_{m+1})=f(u_{1},...,u_{m})=\overline{f}(\overline{u_{1}%
},...,\overline{u_{m}})=\overline{f^{(m+1)}}(\overline{u_{1}},...,\overline
{u_{m+1}})
\]%
\begin{align*}
f_{i\rightarrow j}(u_{1},...,\underset{i}{u_{i}},...,\underset{j}{u_{j}%
},...,u_{m})  &  =f(u_{1},...,\underset{i}{u_{i}},...,\underset{j}{u_{i}%
},...,u_{m})=\\
&  =\overline{f}(\overline{u_{1}},...,\underset{i}{\overline{u_{i}}%
},...,\underset{j}{\overline{u_{i}}},...,\overline{u_{m}})=\overline
{f_{i\rightarrow j}}(\overline{u_{1}},...,\underset{i}{\overline{u_{i}}%
},...,\underset{j}{\overline{u_{j}}},...,\overline{u_{m}})
\end{align*}%
\[
f_{\widehat{u}_{i}}(u_{1},...,\widehat{u}_{i},...,u_{m})=f(u_{1}%
,...,\underset{i}{0},...,u_{m})=\overline{f}(\overline{u_{1}},...,\underset
{i}{\overline{0}},...,\overline{u_{m}})=\overline{f_{\widehat{u}_{i}}%
}(\overline{u_{1}},...,\widehat{\overline{u_{i}}},...,\overline{u_{m}})
\]
and these prove the last three statements of the Theorem.
\end{proof}

\begin{theorem}
If the systems $f,g$ are symmetrical, then the systems $f\cap g$ and $f\cup g
$ are symmetrical.
\end{theorem}

\begin{proof}
$\forall u,\forall x,x\in(f\cap g)(u)\Longleftrightarrow x\in f(u)\quad
and\quad x\in g(u)\Longleftrightarrow\overline{x}\in f(\overline{u})\quad
and\quad\overline{x}\in g(\overline{u})\Longleftrightarrow\overline{x}%
\in(f\cap g)(\overline{u})\Longleftrightarrow x\in\overline{(f\cap
g)}(\overline{u})$

and similarly for the second statement.
\end{proof}

\begin{theorem}
If $f$ is symmetrical, then the next formula is true%
\[
\forall u,\phi(u)=\overline{\phi}(\overline{u})
\]
\end{theorem}

\begin{proof}
$\forall u,\phi(u)=\{x(0-0)|x\in f(u)\}=\{x(0-0)|x\in\overline{f}(\overline
{u})\}=\overline{\phi}(\overline{u})$
\end{proof}

\begin{theorem}
If $f,f^{1},...,f^{p}$ are symmetrical systems, then $f\circ(f^{1},...,f^{p})$
is symmetrical.
\end{theorem}

\begin{proof}
$\forall(u^{1},...,u^{p}),\forall x,x\in f\circ(f^{1},...,f^{p})(u^{1}%
,...,u^{p})\Longleftrightarrow$

$\Longleftrightarrow\exists y^{1}\in f^{1}(u^{1}),...,\exists y^{p}\in
f^{p}(u^{p})\quad s.t.\quad x\in f(y^{1},...,y^{p})$

$\Longleftrightarrow\exists\overline{y^{1}}\in f^{1}(\overline{u^{1}%
}),...,\exists\overline{y^{p}}\in f^{p}(\overline{u^{p}})\quad s.t.\quad
\overline{x}\in f(\overline{y^{1}},...,\overline{y^{p}})$

$\Longleftrightarrow\overline{x}\in f\circ(f^{1},...,f^{p})(\overline{u^{1}%
},...,\overline{u^{p}})\Longleftrightarrow x\in\overline{f\circ(f^{1}%
,...,f^{p})}(\overline{u^{1}},...,\overline{u^{p}})$
\end{proof}

\begin{theorem}
Let $f=X$ an autonomous system, with $X\subset S^{(n)}$. The next statements
are equivalent:

\begin{itemize}
\item[a)] $f$ is symmetrical

\item[b)] $\forall x,x\in X\Longleftrightarrow\overline{x}\in X$
\end{itemize}
\end{theorem}

\begin{proof}
$\forall u,f(u)=f(\overline{u})=X$ and the equivalence between a) and b) is
easily proved
\end{proof}

\begin{corollary}
If $f$ is symmetrical and $X\subset S^{(n)}$ satisfies%
\[
\forall x,x\in X\Longleftrightarrow\overline{x}\in X
\]
then $f\cap X$ and $f\cup X$ are symmetrical.
\end{corollary}

\begin{proof}
From Theorem 84 and 87.
\end{proof}

\bigskip

{\Large 14. Stability}

\begin{definition}
We consider the Boolean function $F:\mathbf{B}^{m}\rightarrow\mathbf{B}^{n}$
and the next properties of the system $f$:

\begin{itemize}
\item[a)] absolute stability%
\[
\forall u,\forall x\in f(u),\exists t_{1},\forall t\geq t_{1},x(t)=x(t_{1})
\]

\item[b)] relative stability%
\[
\forall u,\forall x\in f(u),(\exists t_{1},\forall t\geq t_{1},u(t)=u(t_{1}%
))\Longrightarrow(\exists t_{1},\forall t\geq t_{1},x(t)=x(t_{1}))
\]

\item[c)] stability relative to $F$:%
\[
\forall u,\forall x\in f(u),(\exists t_{1},\forall t\geq t_{1}%
,F(u(t))=F(u(t_{1})))\Longrightarrow(\exists t_{1},\forall t\geq
t_{1},x(t)=x(t_{1}))
\]

\item[d)] delay-insensitivity relative to $F$:%
\[
\forall u,\forall x\in f(u),(\exists t_{1},\forall t\geq t_{1}%
,F(u(t))=F(u(t_{1})))\Longrightarrow(\exists t_{1},\forall t\geq
t_{1},x(t)=F(u(t_{1})))
\]
\end{itemize}
\end{definition}

\begin{remark}
The stability problem is that of the existence of the limit $\underset
{t\rightarrow\infty}{\lim}$ $x(t)$ and Definition 89 states such stability
conditions true for any $u$ and any $x\in f(u)$, the next implications being
true:%
\[%
\begin{array}
[c]{ccccc}%
a) & \Longrightarrow & c) & \Longleftarrow & d)\\
&  & \Downarrow &  & \\
&  & b) &  &
\end{array}
\]
In Definition 89, $F$ is the 'Boolean function to be computed' and $F(u(t))$
is the cause of $x$. When the cause is persistent in the sense that
$\underset{t\rightarrow\infty}{\lim}F(u(t))$ exists and if $f$ is
delay-insensitive relative to $F$, we have $\underset{t\rightarrow\infty}%
{\lim}x(t)=\underset{t\rightarrow\infty}{\lim}F(u(t))$, the so called
'unbounded delay model' giving the manner in which the values of $x$ reproduce
the values of $F(u)$. The stability of $f$ relative to $F$ should be
interpreted like this: when the cause is persistent, thus $\underset
{t\rightarrow\infty}{\lim}F(u(t))$ exists, we have that $\underset
{t\rightarrow\infty}{\lim}x(t)$ exists, thus $f$ is stable, but the two limits
are not necessarily equal; this phenomenon is called hazard when we regard the
states of $f$ as starting, but not completing (correctly) the computation of
$F(u)$ and another possibility exists also that the states of $f(u)$ do not
compute $F(u)$.
\end{remark}

\begin{example}
The systems from Example 13 (9), (10), respectively from Example 63 (11) are
absolutely stable. The system%
\[
f(u)=\{x|\exists t_{1},\forall t\geq t_{1},x(t)=u_{1}(t)\cdot...\cdot
u_{m}(t)\}
\]
is delay-insensitive relative to $F(\lambda)=\lambda_{1}\cdot...\cdot
\lambda_{m}$ and relatively stable, but it is not absolutely stable;%
\[
f(u)=\{x|\exists t_{1},\forall t\geq t_{1},x(t)=%
\genfrac{\{}{.}{0pt}{}{0,\text{ if }\exists\underset{\xi\rightarrow\infty
}{\lim}\text{ }u_{1}(\xi)\cdot...\cdot u_{m}(\xi)}{u_{1}(t),\quad else}%
\}
\]
is relatively stable and stable relative to $F(\lambda)=\lambda_{1}%
\cdot...\cdot\lambda_{m}$ but it is neither absolutely stable, nor
delay-insensitive relative to $F$ and%
\[
f(u)=\{x|\exists t_{1},\forall t\geq t_{1},x(t)=%
\genfrac{\{}{.}{0pt}{}{0,\text{ if }\exists\underset{\xi\rightarrow\infty
}{\lim}\text{ }u(\xi)}{u_{1}(t),\quad else}%
\}
\]
is relatively stable, but it is not absolutely stable. For $F(\lambda
)=\lambda_{2}$ , by taking $(u_{1},u_{2},...,u_{m})=(\chi_{\lbrack
0,1)\cup\lbrack2,3)\cup\lbrack4,5)\cup...},0,...,0)$ we remark that $f$ is not
stable relative to $F$.
\end{example}

\begin{theorem}
The next statements are equivalent:

a) $f$ is absolutely stable

b) $f$ is stable relative to the constant function

and the next statements are also equivalent for $\mu\in\mathbf{B}^{n}$:

i) $\forall u,\forall x\in f(u),\exists t_{1},\forall t\geq t_{1},x(t)=\mu$

ii) $f$ is delay-insensitive relative to the constant function $F=\mu$.
\end{theorem}

\begin{proof}
a)$\Longleftrightarrow$b) is true because a) is the conclusion of b), where b)
has a hypothesis always fulfilled.

i)$\Longleftrightarrow$ii) takes place in similar conditions with the previous equivalence.
\end{proof}

\begin{theorem}
If $F,G:\mathbf{B}^{m}\rightarrow\mathbf{B}^{n}$ are two Boolean functions
with%
\begin{equation}
\forall\lambda,\forall\lambda^{\prime},F(\lambda)=F(\lambda^{\prime
})\Longrightarrow G(\lambda)=G(\lambda^{\prime})
\end{equation}
and if the system $f$ is stable relative to $G$, then it is stable relative to
$F$.
\end{theorem}

\begin{proof}
We suppose that $f$ is stable relative to $G$:%
\[
\forall u,\forall x\in f(u),(\exists t_{1},\forall t\geq t_{1}%
,G(u(t))=G(u(t_{1})))\Longrightarrow(\exists t_{1},\forall t\geq
t_{1},x(t)=x(t_{1}))
\]
and let $u,x\in f(u)$ arbitrary so that%
\[
\exists t_{1},\forall t\geq t_{1},F(u(t))=F(u(t_{1}))
\]
The hypothesis (12) states that%
\[
\exists t_{1},\forall t\geq t_{1},G(u(t))=G(u(t_{1}))
\]
from where%
\[
\exists t_{1},\forall t\geq t_{1},x(t)=x(t_{1})
\]
and $f$ is stable relative to $F$.
\end{proof}

\begin{theorem}
Let the Boolean function $F$ and the system $f$. If $f$ is absolutely stable
(relatively stable, stable relative to $F$, delay-insensitive relative to
$F$), then the systems $\overline{f},f^{(m+1)},f_{i\rightarrow j}%
,f_{\widehat{u}_{i}}$ are absolutely stable (relatively stable, stable
relative to $\overline{F},F^{(m+1)},F_{i\rightarrow j},F_{\widehat{u}_{i}}$,
delay-insensitive relative to $\overline{F},F^{(m+1)},F_{i\rightarrow
j},F_{\widehat{u}_{i}}$), where $i,j\in\{1,...,m\},i\neq j$ and $\overline
{F},F^{(m+1)},F_{i\rightarrow j},F_{\widehat{u}_{i}}$ are defined by:%
\[
\overline{F}:\mathbf{B}^{m}\rightarrow\mathbf{B}^{n},\overline{F}(\lambda
_{1},...,\lambda_{m})=\overline{F(\lambda_{1},...,\lambda_{m})}%
\]%
\[
F^{(m+1)}:\mathbf{B}^{m+1}\rightarrow\mathbf{B}^{n},F^{(m+1)}(\lambda
_{1},...,\lambda_{m+1})=F(\lambda_{1},...,\lambda_{m})
\]%
\[
F_{i\rightarrow j}:\mathbf{B}^{m}\rightarrow\mathbf{B}^{n},F_{i\rightarrow
j}(\lambda_{1},...,\lambda_{m})=F(\lambda_{1},...,\underset{i}{\lambda_{i}%
},...,\underset{j}{\lambda_{i}},...,\lambda_{m})
\]%
\[
F_{\widehat{u}_{i}}:\mathbf{B}^{m-1}\rightarrow\mathbf{B}^{n},F_{\widehat
{u}_{i}}(\lambda_{1},...,\widehat{\lambda}_{i},...,\lambda_{m})=F(\lambda
_{1},...,\underset{i}{0},...,\lambda_{m})
\]
\end{theorem}

\begin{proof}
We suppose that $f$ is delay insensitive relative to $F$:%
\[
\forall u,\forall x\in f(u),(\exists t_{1},\forall t\geq t_{1}%
,F(u(t))=F(u(t_{1})))\Longrightarrow(\exists t_{1},\forall t\geq
t_{1},x(t)=F(u(t_{1})))
\]
from where%
\[
\forall u,\forall\overline{x}\in f(u),(\exists t_{1},\forall t\geq
t_{1},\overline{F}(u(t))=\overline{F}(u(t_{1})))\Longrightarrow(\exists
t_{1},\forall t\geq t_{1},\overline{x}(t)=\overline{F}(u(t_{1})))
\]
i.e. $\overline{f}$ is delay-insensitive relative to $\overline{F}$. Moreover,
we observe that%
\[
\forall(u_{1},...,u_{m+1}),\forall x\in f(u_{1},...,u_{m})=f^{(m+1)}%
(u_{1},...,u_{m+1}),
\]%
\[
(\exists t_{1},\forall t\geq t_{1},F^{(m+1)}(u_{1}(t),...,u_{m+1}%
(t))=F^{(m+1)}(u_{1}(t_{1}),...,u_{m+1}(t_{1})))\Longleftrightarrow
\]%
\[
\Longleftrightarrow(\exists t_{1},\forall t\geq t_{1},F(u_{1}(t),...,u_{m}%
(t))=F(u_{1}(t_{1}),...,u_{m}(t_{1})))\Longrightarrow
\]%
\[
\Longrightarrow(\exists t_{1},\forall t\geq t_{1},x(t)=F(u_{1}(t_{1}%
),...,u_{m}(t_{1}))\Longleftrightarrow
\]%
\[
\Longleftrightarrow(\exists t_{1},\forall t\geq t_{1},x(t)=F^{(m+1)}%
(u_{1}(t_{1}),...,u_{m+1}(t_{1}))
\]
meaning that $f^{(m+1)}$ is delay-insensitive relative to $F^{(m+1)}$ etc
\end{proof}

\begin{remark}
If $f$ is stable relative to $F$, then it is stable relative to $\overline{F}$.
\end{remark}

\begin{theorem}
Let the systen $f$ be absolutely stable (relatively stable, stable relative to
$F$, delay-insensitive relative to $F$). The next statements are true:

\begin{itemize}
\item[a)] Any system $f^{\prime}\subset f$ is absolutely stable (relatively
stable, stable relative to $F$, delay-insensitive relative to $F$)

\item[b)] If the system $g$ is absolutely stable (relatively stable, stable
relative to $F$, delay-insensitive relative to $F$) then $f\cup g$ is
absolutely stable (relatively stable, stable relative to $F$,
delay-insensitive relative to $F$).
\end{itemize}
\end{theorem}

\begin{proof}
$b)$ We suppose that $f,g$ are delay-insensitive relative to $F$ and let
$u,x\in(f\cup g)(u)$ arbitrary, for example $x\in f(u)$. We have%
\[
(\exists t_{1},\forall t\geq t_{1},F(u(t))=F(u(t_{1})))\Longrightarrow(\exists
t_{1},\forall t\geq t_{1},x(t)=F(u(t_{1})))
\]
from where we infer the delay-insensitivity of $f\cup g$ relative to $F$.
\end{proof}

\begin{corollary}
If $f$ is absolutely stable (relatively stable, stable relative to $F$,
delay-insensitive relative to $F$), then $f\cap X$ and $f\cap g$ are
absolutely stable (relatively stable, stable relative to $F$,
delay-insensitive relative to $F$), for any $X\subset S^{(n)}$ and any system
$g$.
\end{corollary}

\begin{proof}
Special case of Theorem 96 $a)$.
\end{proof}

\begin{theorem}
Let the functions $F:\mathbf{B}^{m}\rightarrow\mathbf{B}^{n},F^{i}%
:\mathbf{B}^{m_{i}}\rightarrow\mathbf{B}^{n_{i}},i=\overline{1,p}$ \ and the
systems $f:S^{(m)}\rightarrow P^{\ast}(S^{(n)}),f^{i}:S^{(m_{i})}\rightarrow
P^{\ast}(S^{(n_{i})}),i=\overline{1,p}$ so that $n_{1}+...+n_{p}=m$ .

\begin{itemize}
\item[a)] If $f,f^{1},...,f^{p}$ are relatively stable (stable relative to
$F,F^{1},...,F^{p}$, delay-insensitive relative to $F,F^{1},...,F^{p}$), then
$f\circ(f^{1},...,f^{p})$ is relatively stable (stable relative to
$F\circ(F^{1},...,F^{p})$, delay-insensitive relative to $F\circ
(F^{1},...,F^{p})$)

\item[b)] If $f$ is absolutely stable, then $f\circ(f^{1},...,f^{p})$ is
absolutely stable.
\end{itemize}
\end{theorem}

\begin{proof}
a) We suppose for example that $f^{1},...,f^{p}$ are stable relative to
$F^{1},...,F^{p}$:%
\[
\forall u^{1},\forall y^{1}\in f^{1}(u^{1}),(\exists t_{1},\forall t\geq
t_{1},F^{1}(u^{1}(t))=F^{1}(u^{1}(t_{1})))\Longrightarrow(\exists
t_{1},\forall t\geq t_{1},y^{1}(t)=y^{1}(t_{1}))
\]%
\[
...
\]%
\[
\forall u^{p},\forall y^{p}\in f^{p}(u^{p}),(\exists t_{1},\forall t\geq
t_{1},F^{p}(u^{p}(t))=F^{p}(u^{p}(t_{1})))\Longrightarrow(\exists
t_{1},\forall t\geq t_{1},y^{p}(t)=y^{p}(t_{1}))
\]
or equivalently%
\begin{align*}
\forall(u^{1},...,u^{p}),\forall y  &  \in(f^{1},...,f^{p})(u^{1}%
,...,u^{p}),\\
(\exists t_{1},\forall t  &  \geq t_{1},(F^{1},...,F^{p})(u^{1}(t),...,u^{p}%
(t))=(F^{1},...,F^{p})(u^{1}(t_{1}),...,u^{p}(t_{1})))\Longrightarrow\\
&  \Longrightarrow(\exists t_{1},\forall t\geq t_{1},y(t)=y(t_{1}))
\end{align*}
We have noted $y=(y^{1},...,y^{p})$ and we suppose from now that
$(u^{1},...,u^{p}),y$ are arbitrary, fixed. Because $f$ is stable relative to
$F$, we can write%
\[
\forall x\in f(y),(\exists t_{1},\forall t\geq t_{1},F(y(t))=F(y(t_{1}%
)))\Longrightarrow(\exists t_{1},\forall t\geq t_{1},x(t)=x(t_{1}))
\]
$f\circ(f^{1},...,f^{p})$ is stable relative to $F\circ(F^{1},...,F^{p})$.

b) Let $u^{1},...,u^{p},y^{1}\in f^{1}(u^{1}),...,y^{p}\in f^{p}(u^{p})$ and
$x\in f(y^{1},...,y^{p})$ arbitrary. $t_{1}$ exists so that $\forall t\geq
t_{1},x(t)=x(t_{1})$ from where the conclusion that $f\circ(f^{1},...,f^{p})$
is absolutely stable follows.
\end{proof}

\begin{theorem}
Let the Boolean function $F$ and the set $X\subset S^{(n)}$ that is identified
with an autonomous system $f$. The next statements are equivalent:

a) $\forall x\in X,\exists t_{1},\forall t\geq t_{1},x(t)=x(t_{1})$

b) $f$ is absolutely stable

c) $f$ is relatively stable

d) $f$ is stable relative to $F$

and the next statements are also equivalent for some $\mu\in\mathbf{B}^{n}$:

i) $\forall x\in X,\exists t_{1},\forall t\geq t_{1},x(t)=\mu$

ii) $f$ is delay-insensitive relative to the constant function $F=\mu$.
\end{theorem}

\begin{proof}
a) and b) are obviously equivalent. We suppose that $f$ is relatively stable
and we choose $u$ so that $\exists t_{1},\forall t\geq t_{1},u(t)=u(t_{1})$.
Then a) takes place and because the hypothesis depending on $u$ implies a
conclusion that is independent on $u$, we have that c) implies a). The
implication a)$\Longrightarrow$c) is obvious.

a)$\Longleftrightarrow$d) is shown similarly with a)$\Longleftrightarrow$c).

i)$\Longleftrightarrow$ii) takes place because i) is the conclusion of the
request of delay-insensitivity of $f$ relative to $F=\mu$.
\end{proof}

\bigskip

{\Large 15. Fundamental mode}

\begin{definition}
Let the Boolean function $F:\mathbf{B}^{m}\rightarrow\mathbf{B}^{n}$, the
system $f:S^{(m)}\rightarrow P^{\ast}(S^{(n)})$, the input $u\in S^{(m)}$ and
the state $x\in f(u)$. We suppose that an unbounded sequence $0\leq
t_{0}<t_{1}<t_{2}<...$ exists so that the next properties be stated:%
\begin{equation}
\forall t<t_{0},u(t)=u(t_{0}-0)
\end{equation}%
\begin{equation}
\forall k\geq0,\forall t\in\lbrack t_{k},t_{k+1}),u(t)=u(t_{k})
\end{equation}%
\begin{equation}
\forall k\geq0,\forall t\in\lbrack t_{k},t_{k+1}),F(u(t))=F(u(t_{k}))
\end{equation}%
\begin{equation}
\forall k\geq1,x\cdot\chi_{(-\infty,t_{k})}\oplus x(t_{k}-0)\cdot\chi_{\lbrack
t_{k},\infty)}\in f(u\cdot\chi_{(-\infty,t_{k})}\oplus u(t_{k}-0)\cdot
\chi_{\lbrack t_{k},\infty)})
\end{equation}%
\begin{equation}
\forall k\geq1,x(t_{k}-0)=F(u(t_{k}-0))
\end{equation}
The couple $(u,x)$ is called

a) a pseudo-fundamental (operating) mode of $f$ if (16) is true

b) a fundamental (operating) mode of $f$ if (13), (14), (16) are true

c) a fundamental (operating) mode of $f$ relative to $F$ if (13), (15), (16)
are true

d) a delay-insensitive fundamental (operating) mode of $f$ relative to $F$ if
(13), (15), (16), (17) are true.
\end{definition}

\begin{remark}
$(u,x)$ is a pseudo-fundamental mode of $f$ if the intervals $[t_{k-1},t_{k})$
covering $[0,\infty)$ exist (from the unboundness of $t_{0},t_{1},t_{2},...$)
having the property that $u$ is allowed to take new values in $[t_{k}%
,t_{k+1})$ possibly different from the previous ones in $[t_{k-1},t_{k})$ only
if $x$ has stabilized (at some time instant situated in the interval
$[t_{k-1},t_{k})$ ) under the input $u\cdot\chi_{(-\infty,t_{k})}\oplus
u(t_{k}-0)\cdot\chi_{\lbrack t_{k},\infty)}$ to the value $x(t_{k}-0)$. The
forms of $u,x$ do not matter, just the satisfaction of the stability condition
100 (16), that characterizes $t_{1},t_{2},t_{3},...$ as time instants when
$u,x$ are in equilibrium. We shall consider that $(u,x)$ are in equilibrium in
$t_{0}$ too.

$(u,x)$ is a fundamental mode of $f$ if the satisfaction of the stability
condition 100 (16) takes place for $u$ constant in $(-\infty,t_{0})$ and also
in each interval $[t_{k-1},t_{k})$ and $(u,x)$ is a fundamental mode of $f$
relative to $F$ if the condition 100 (14) is relaxed to 100 (15). Here the
role of $F$ is that of 'Boolean function to be computed', to be compared with
the delay-insensitivity of $f$ relative to $F$, Definition 89 d), whose
hypothesis $\exists t_{1},\forall t\geq t_{1},F(u(t))=F(u(t_{1}))$ was
replaced by 100 (15) and whose conclusion $\exists t_{1},\forall t\geq
t_{1},x(t)=F(u(t_{1}))$ was replaced by 100 (16), (17).

The absence of the satisfaction of 100 (17) between the previous properties
indicates either the presence of hazard: the states of the system are supposed
to start the computation of $F(u)$ and this computation is unsuccessful
eventually, or the fact that the state $x\in f(u)$ is not related with the
computation of $F(u)$.

The definitions that are grouped in 100 include the possibility $u(t_{k}%
)=u(t_{k+1})$, respectively $F(u(t_{k}))=F(u(t_{k+1}))$ or $\exists
k,u(t_{k})=u(t_{k+1})=...$, respectively $\exists k,F(u(t_{k}))=F(u(t_{k+1}))=...$
\end{remark}

\begin{theorem}
For $F,f,u,x\in f(u)$ and $0\leq t_{0}<t_{1}<t_{2}<...$ unbounded, we suppose
that some of the requests 100 (13),...,(17) are satisfied. The same properties
are satisfied if we replace the sequence $0\leq t_{0}<t_{1}<t_{2}<...$ \ with
$0\leq t_{0}^{^{\prime}}<t_{1}^{^{\prime}}<t_{2}^{^{\prime}}<...$ \ where%
\[
t_{0}^{^{\prime}}=t_{0},t_{1}^{^{\prime}}=t_{1},...,t_{k}^{^{\prime}}%
=\tau,t_{k+1}^{^{\prime}}=t_{k},t_{k+2}^{^{\prime}}=t_{k+1},...
\]
with $k\geq1$ arbitrary and $\tau\in(t_{k-1},t_{k})$ chosen sufficiently close
to $t_{k}$.
\end{theorem}

\begin{proof}
We fix $k\geq1$ and $t_{k}$ arbitrary the next properties that derive from 100
(14), (16), (17) being satisfied%
\begin{equation}
\forall t\in\lbrack t_{k-1},t_{k}),u(t)=u(t_{k-1})
\end{equation}%
\begin{equation}
x\cdot\chi_{(-\infty,t_{k})}\oplus x(t_{k}-0)\cdot\chi_{\lbrack t_{k},\infty
)}\in f(u\cdot\chi_{(-\infty,t_{k})}\oplus u(t_{k}-0)\cdot\chi_{\lbrack
t_{k},\infty)})
\end{equation}%
\begin{equation}
x(t_{k}-0)=F(u(t_{k}-0))
\end{equation}
We have the existence of some $\varepsilon>0$ so that%
\[
\forall t\in(t_{k}-\varepsilon,t_{k}),x(t)=x(t_{k}-0)
\]
because $x$ has a left limit in $t_{k}$ and for any $\tau\in(t_{k}%
-\varepsilon,t_{k})\cap(t_{k-1},t_{k})$ we infer the truth of%
\[
\forall t\in\lbrack t_{k-1},\tau),u(t)=u(t_{k-1})\quad and\quad\forall
t\in\lbrack\tau,t_{k}),u(t)=u(\tau)
\]%
\[
x\cdot\chi_{(-\infty,\tau)}\oplus x(\tau-0)\cdot\chi_{\lbrack\tau,\infty)}\in
f(u\cdot\chi_{(-\infty,\tau)}\oplus u(\tau-0)\cdot\chi_{\lbrack\tau,\infty
)})\quad and\quad(19)
\]%
\[
x(\tau-0)=F(u(\tau-0))\quad and\quad(20)
\]
thus the insertion of such a $\tau$ between the elements of $(t_{k})$ leaves
the relations 100 (14), (16), (17) true. The situation is similar if we refer
to 100 (15) instead of 100 (14).
\end{proof}

\begin{definition}
Let $(u,x)$ a pseudo-fundamental mode of $f$ (a fundamental mode of $f$, a
fundamental mode of $f$ relative to $F$, a delay-insensitive fundamental mode
of $f$ relative to $F$) and the unbounded sequence $0\leq t_{0}<t_{1}%
<t_{2}<...$ with the property that the relations 100 (16) (the relations 100
(13), (14), (16), the relations 100 (13), (15), (16), the relations 100 (13),
(15), (16), (17)) are fulfilled. Then we say that the sequence $(t_{k})$ is
compatible with the mode $(u,x)$.
\end{definition}

\begin{definition}
We suppose that $(u,x)$ is a pseudo-fundamental mode of $f$ and let $0\leq
t_{0}<t_{1}<t_{2}<...$ compatible with it. The functions%
\begin{equation}
u^{(k)}=u\cdot\chi_{(-\infty,t_{k})}\oplus u(t_{k}-0)\cdot\chi_{\lbrack
t_{k},\infty)}%
\end{equation}%
\begin{equation}
x^{(k)}=x\cdot\chi_{(-\infty,t_{k})}\oplus x(t_{k}-0)\cdot\chi_{\lbrack
t_{k},\infty)}%
\end{equation}
$k\geq1$ are called initial segments, or prefixes (relative to $(t_{k})$) of
$u,x$ and the couples $(u(t_{k}-0),x(t_{k}-0)),k\geq1$ are called points of
equilibrium of $f$. By definition $(u(t_{0}-0),x(t_{0}-0))$ is a point of
equilibrium of $f$ too.
\end{definition}

\begin{theorem}
$F,f,u,x\in f(u)$ and the unbounded sequence $0\leq t_{0}<t_{1}<t_{2}<...$ are given.

a) Let $(u,x)$ a pseudo-fundamental mode of $f$ (a fundamental mode of $f$, a
fundamental mode of $f$ relative to $F$, a delay-insensitive fundamental mode
of $f$ relative to $F$) so that $(t_{k})$ be compatible with it. Then
$(u^{(k)},$ $x^{(k)})$ are pseudo-fundamental modes of $f$ (fundamental modes
of $f$, fundamental modes of $f$ relative to $F$, delay-insensitive
fundamental modes of $f$ relative to $F$) for all $k\geq1$.

b) Let the couples $(u\cdot\chi_{(-\infty,t_{k})}\oplus u(t_{k}-0)\cdot
\chi_{\lbrack t_{k},\infty)},$ $x\cdot\chi_{(-\infty,t_{k})}\oplus
x(t_{k}-0)\cdot\chi_{\lbrack t_{k},\infty)})$ pseudo-fundamental modes of $f$
(fundamental modes of $f$, fundamental modes of $f$ relative to $F$,
delay-insensitive fundamental modes of $f$ relative to $F$) for all $k\geq1$.
Then $(u,x)$ is a pseudo-fundamental mode of $f$ (a fundamental mode of $f$, a
fundamental mode of $f$ relative to $F$, a delay-insensitive fundamental mode
of $f$ relative to $F$) and $(t_{k})$ is compatible with it.
\end{theorem}

\begin{proof}
a) We suppose for example that 100 (13), (15), (16), (17) are satisfied, we
fix $k^{\prime}\geq1$ and we infer%
\[
\forall t<t_{0},u^{(k^{\prime})}(t)=u^{(k^{\prime})}(t_{0}-0)=u(t_{0}-0)
\]%
\[
\forall t\in\lbrack t_{k},t_{k+1}),F(u^{(k^{\prime})}(t))=F(u^{(k^{\prime}%
)}(t_{k}))=%
\genfrac{\{}{.}{0pt}{}{F(u(t_{k})),0\leq k<k^{\prime}}{F(u(t_{k^{\prime}%
-1})),k\geq k^{\prime}}%
\]
On the other hand, the property%
\[
x^{(k^{\prime})}\cdot\chi_{(-\infty,t_{k})}\oplus x^{(k^{\prime})}%
(t_{k}-0)\cdot\chi_{\lbrack t_{k},\infty)}\in f(u^{(k^{\prime})}\cdot
\chi_{(-\infty,t_{k})}\oplus u^{(k^{\prime})}(t_{k}-0)\cdot\chi_{\lbrack
t_{k},\infty)})
\]
coincides with 100 (16) for $1\leq k\leq k^{\prime}$ and with%
\[
x^{(k^{\prime})}\cdot\chi_{(-\infty,t_{k^{\prime}})}\oplus x^{(k^{\prime}%
)}(t_{k^{\prime}}-0)\cdot\chi_{\lbrack t_{k^{\prime}},\infty)}\in
f(u^{(k^{\prime})}\cdot\chi_{(-\infty,t_{k^{\prime}})}\oplus u^{(k^{\prime}%
)}(t_{k^{\prime}}-0)\cdot\chi_{\lbrack t_{k^{\prime}},\infty)})
\]
for $k>k^{\prime}$ and eventually the property%
\[
x^{(k^{\prime})}(t_{k}-0)=F(u^{(k^{\prime})}(t_{k}-0))
\]
coincides with 100 (17) for $1\leq k\leq k^{\prime}$ and with%
\[
x^{(k^{\prime})}(t_{k^{\prime}}-0)=F(u^{(k^{\prime})}(t_{k^{\prime}}-0))
\]
for $k>k^{\prime}$. $(u^{(k^{\prime})},x^{(k^{\prime})})$ is a delay
insensitive fundamental mode of $f$ relative to $F$, the property being true
for any $k^{\prime}\geq1$.

b) Let $u,x\in f(u),0\leq t_{0}<t_{1}<t_{2}<...$ unbounded and $k^{\prime}%
\geq1$ arbitrary, fixed so that $u^{(k^{\prime})},x^{(k^{\prime})}$ defined
like at 104 (21), (22) satisfy for example 100 (13), (15), (16), (17) i.e.
$(u^{(k^{\prime})},x^{(k^{\prime})})$ is a delay-insensitive fundamental mode
of $f$ relative to $F$. $u,x$ satisfy 100 (13); 100 (15), (16), (17) are
satisfied for $0\leq k\leq k^{\prime},$ $1\leq k\leq k^{\prime}$, $1\leq k\leq
k^{\prime}$ and when $k^{\prime}$ is variable, we have that $(u,x)$ is a
delay-insensitive fundamental mode of $f$ relative to $F$.
\end{proof}

\begin{theorem}
\bigskip Let $F,f,u$ and $x\in f(u)$. The next statements are true:

a) If $(u,x)$ is a fundamental mode of $f$, then $(u,x)$ is a fundamental mode
of $f$ relative to $F$

b) If $F$ is injective and $(u,x)$ is a fundamental mode of $f$ relative to
$F$, then $(u,x)$ is a fundamental mode of $f$

c) If $(u,x)$ is a fundamental mode of $f$ (relative to $F$), then it is a
pseudo-fundamental mode of $f.$
\end{theorem}

\begin{proof}
100 (14) implies 100 (15) for any $F$ and if $F$ is injective, then 100 (15)
implies 100 (14).
\end{proof}

\begin{theorem}
The next statements are equivalent:

a) $(u,x)$ is a fundamental mode of $f$

b) for any function $F$, $(u,x)$ is a fundamental mode of $f$ relative to $F.$
\end{theorem}

\begin{proof}
$b)\Longrightarrow a)$ Let $F^{i}:\mathbf{B}^{m}\rightarrow\mathbf{B}%
^{n},\forall\lambda\in\mathbf{B}^{m},F^{i}(\lambda)=(\lambda_{i},0,...,0)$ and
$0\leq t_{0}^{i}<t_{1}^{i}<t_{2}^{i}<...$ unbounded so that 100 (13), (15),
(16) be satisfied for all $i\in\{1,...,m\}$. If $0\leq t_{0}<t_{1}<t_{2}<...$
is the sequence obtained by indexing the family $(t_{k}^{1})\cup...\cup
(t_{k}^{m})$ we remark that 100 (13), (14), (16) are fulfilled.
\end{proof}

\begin{theorem}
a) Let the non-anticipatory (Definition 50) relatively stable system $f$ and
the family of vectors $u^{k}\in\mathbf{B}^{m},k\in\mathbf{N}$. The input $u\in
S^{(m)}$:%
\[
u(t)=u^{0}\cdot\chi_{\lbrack t_{o},t_{1})}(t)\oplus u^{1}\cdot\chi_{\lbrack
t_{1},t_{2})}(t)\oplus...
\]
and the state $x\in f(u)$ exist so that $(u,x)$ is a fundamental mode of $f$.

b) Let the non-anticipatory (Definition 50) relatively stable systems
$f,f^{1},...,f^{p}$ and the family of vectors $z^{k}\in\mathbf{B}%
^{m_{1}+...+m_{p}},k\in\mathbf{N}$. The input $z\in S^{(m_{1}+...+m_{p})}$:%
\[
z(t)=z^{0}\cdot\chi_{\lbrack t_{o},t_{1})}(t)\oplus z^{1}\cdot\chi_{\lbrack
t_{1},t_{2})}(t)\oplus...
\]
and the state $x\in f\circ(f^{1},...,f^{p})(z)$ exist so that $(z,x)$ is a
fundamental mode of $f\circ(f^{1},...,f^{p})$.
\end{theorem}

\begin{proof}
b) We consider the family of vectors $z^{k}\in\mathbf{B}^{m_{1}+...+m_{p}%
},k\in\mathbf{N}$ and we fix $t_{0}\geq0$ arbitrary. For the input%
\[
z^{(1)}(t)=z^{0}\cdot\chi_{\lbrack t_{0},\infty)}(t)
\]
from the relative stability of $f,f^{1},...,f^{p}$ we infer the existence of
$y^{(1)}\in(f^{1},...,f^{p})(z^{(1)}),$ $x^{(1)}\in f(y^{(1)})$ and
$t_{1}>t_{0}$ so that%
\[
y^{(1)}(t)=y^{(1)}(t)\cdot\chi_{(-\infty,t_{1})}(t)\oplus y^{(1)}%
(t_{1}-0)\cdot\chi_{\lbrack t_{1},\infty)}(t)
\]%
\[
x^{(1)}(t)=x^{(1)}(t)\cdot\chi_{(-\infty,t_{1})}(t)\oplus x^{(1)}%
(t_{1}-0)\cdot\chi_{\lbrack t_{1},\infty)}(t)
\]
We define%
\[
z^{(2)}(t)=z^{0}\cdot\chi_{\lbrack t_{0},t_{1})}(t)\oplus z^{1}\cdot
\chi_{\lbrack t_{1},\infty)}(t)
\]
From the non-anticipation and the relative stability of $f,f^{1},...,f^{p}$ we
infer the existence of $y^{(2)}\in(f^{1},...,f^{p})(z^{(2)}),$ $x^{(2)}\in
f(y^{(2)})$ and $t_{2}>t_{1}$ so that%
\[
y^{(2)}(t)=y^{(1)}(t)\cdot\chi_{(-\infty,t_{1})}(t)\oplus y^{(2)}(t)\cdot
\chi_{\lbrack t_{1},t_{2})}(t)\oplus y^{(2)}(t_{2}-0)\cdot\chi_{\lbrack
t_{2},\infty)}(t)
\]%
\[
x^{(2)}(t)=x^{(1)}(t)\cdot\chi_{(-\infty,t_{1})}(t)\oplus x^{(2)}(t)\cdot
\chi_{\lbrack t_{1},t_{2})}(t)\oplus x^{(2)}(t_{2}-0)\cdot\chi_{\lbrack
t_{2},\infty)}(t)
\]
We can define in this moment%
\[
z^{(3)}(t)=z^{0}\cdot\chi_{\lbrack t_{0},t_{1})}(t)\oplus z^{1}\cdot
\chi_{\lbrack t_{1},t_{2})}(t)\oplus z^{2}\cdot\chi_{\lbrack t_{2},\infty
)}(t)
\]%
\[
...
\]
By using iteratively the non-anticipation and the relative stability of
$f,f^{1},...,f^{p}$ we obtain%
\[
z^{(k+1)}(t)=z^{0}\cdot\chi_{\lbrack t_{0},t_{1})}(t)\oplus z^{1}\cdot
\chi_{\lbrack t_{1},t_{2})}(t)\oplus...\oplus z^{k}\cdot\chi_{\lbrack
t_{k},\infty)}(t)
\]
$y^{(k+1)}\in(f^{1},...,f^{p})(z^{(k+1)}),$ $x^{(k+1)}\in f(y^{(k+1)})$ and
$t_{k+1}>t_{k}$ so that%
\[
y^{(k+1)}(t)=y^{(1)}(t)\cdot\chi_{(-\infty,t_{1})}(t)\oplus y^{(2)}%
(t)\cdot\chi_{\lbrack t_{1},t_{2})}(t)\oplus...\oplus y^{(k+1)}(t_{k+1}%
-0)\cdot\chi_{\lbrack t_{k+1},\infty)}(t)
\]%
\[
x^{(k+1)}(t)=x^{(1)}(t)\cdot\chi_{(-\infty,t_{1})}(t)\oplus x^{(2)}%
(t)\cdot\chi_{\lbrack t_{1},t_{2})}(t)\oplus...\oplus x^{(k+1)}(t_{l+1}%
-0)\cdot\chi_{\lbrack t_{k+1},\infty)}(t)
\]
The functions%
\[
z(t)=z^{0}\cdot\chi_{\lbrack t_{0},t_{1})}(t)\oplus z^{1}\cdot\chi_{\lbrack
t_{1},t_{2})}(t)\oplus...
\]%
\[
x(t)=x^{(1)}\cdot\chi_{(-\infty,t_{1})}(t)\oplus x^{(2)}\cdot\chi_{\lbrack
t_{1},t_{2})}(t)\oplus...
\]
satisfy the required property.
\end{proof}

\begin{theorem}
a) Let the Boolean function $F$, the family of vectors $x^{k}\in Range(F),$
$k\in\mathbf{N}$ and the non-anticipatory (Definition 50) system $f$ that is
stable relative to $F$ (that is delay-insensitive relative to $F$). The input
$u\in S^{(m)}$:%
\[
u(t)=u^{0}\cdot\chi_{\lbrack t_{o},t_{1})}(t)\oplus u^{1}\cdot\chi_{\lbrack
t_{1},t_{2})}(t)\oplus...
\]
and the state $x\in f(u)$ exist so that%
\[
\forall k\in\mathbf{N},F(u^{k})=x^{k}%
\]
and $(u,x)$ is a fundamental mode of $f$ relative to $F$ (a delay-insensitive
fundamental mode of $f$ relative to $F$).

b) Let the Boolean functions $F,F^{1},...,F^{p}$, the family of vectors
$x^{k}\in Range(F\circ(F^{1},...,F^{p})),$ $k\in\mathbf{N}$ and the
non-anticipatory (Definition 50) systems $f,f^{1},...,f^{p}$ that are stable
relative to $F,F^{1},...,F^{p}$ (that are delay-insensitive relative to
$F,F^{1},...,F^{p}$). The input $z\in S^{(m_{1}+...+m_{p})}$:%
\[
z(t)=z^{0}\cdot\chi_{\lbrack t_{o},t_{1})}(t)\oplus z^{1}\cdot\chi_{\lbrack
t_{1},t_{2})}(t)\oplus...
\]
and the state $x\in f\circ(f^{1},...,f^{p})(z)$ exist so that%
\[
\forall k\in\mathbf{N},F\circ(F^{1},...,F^{p})(z^{k})=x^{k}%
\]
and $(z,x)$ is a fundamental mode of $f\circ(f^{1},...,f^{p})$ relative to
$F\circ(F^{1},...,F^{p})$ (a delay-insensitive fundamental mode of
$f\circ(f^{1},...,f^{p})$ relative to $F\circ(F^{1},...,F^{p})$).
\end{theorem}

\begin{proof}
b) We choose arbitrarily the family $z^{k}\in\mathbf{B}^{m_{1}+...+m_{p}}$ so
that $x^{k}=F\circ(F^{1},...,F^{p})(z^{k})$,$k\in\mathbf{N}$ and the proof
coincides with the one from 108 b), where 'relative stability' is replaced by
'stability relative to $F,F^{1},...,F^{p}$'. We have in addition the condition
of delay-insensitivity stating%
\[
y^{(k)}(t_{k}-0)=(F^{1},...,F^{p})(z^{(k)}(t_{k}-0))
\]%
\[
x^{(k)}(t_{k}-0)=F(y^{(k)}(t_{k}-0))
\]
for all $k\geq1$, from where we get%
\[
\forall k\geq1,x(t_{k}-0)=x^{(k)}(t_{k}-0)=F(y^{(k)}(t_{k}-0))=
\]%
\[
=F\circ(F^{1},...,F^{p})(z^{(k)}(t_{k}-0))=F\circ(F^{1},...,F^{p}%
)(z(t_{k}-0))
\]
\end{proof}

\begin{theorem}
We suppose that $(u,x)$ is a pseudo-fundamental mode of $f$ (a fundamental
mode of $f$, a fundamental mode of $f$ relative to $F$, a delay-insensitive
fundamental mode of $f$ relative to $F$). Then $(u,\overline{x})$ is a
pseudo-fundamental mode of $\overline{f}$ (a fundamental mode of $\overline
{f}$, a fundamental mode of $\overline{f}$ relative to $\overline{F}$, a
delay-insensitive fundamental mode of $\overline{f}$ relative to $\overline
{F}$).
\end{theorem}

\begin{proof}
Equations 100 (15),...,(17) imply%
\[
\forall k\geq0,\forall t\in\lbrack t_{k},t_{k+1}),\overline{F}(u(t))=\overline
{F}(u(t_{k}))
\]%
\[
\forall k\geq1,\overline{x}\cdot\chi_{(-\infty,t_{k})}\oplus\overline{x}%
(t_{k}-0)\cdot\chi_{\lbrack t_{k},\infty)}\in\overline{f}(u\cdot\chi
_{(-\infty,t_{k})}\oplus u(t_{k}-0)\cdot\chi_{\lbrack t_{k},\infty)})
\]%
\[
\forall k\geq1,\overline{x}(t_{k}-0)=\overline{F}(u(t_{k}-0))
\]
showing the statements of the Theorem.
\end{proof}

\begin{theorem}
If $(u,x)$ is a pseudo-fundamental mode of $f$ (a fundamental mode of $f$, a
fundamental mode of $f$ relative to $F$, a delay-insensitive fundamental mode
of $f$ relative to $F$) and $f\subset g$, then $(u,x)$ is a pseudo-fundamental
mode of $g$ (a fundamental mode of $g$, a fundamental mode of $g$ relative to
$F$, a delay-insensitive fundamental mode of $g$ relative to $F$).
\end{theorem}

\begin{proof}
The condition%
\[
\forall k\geq1,x\cdot\chi_{(-\infty,t_{k})}\oplus x(t_{k}-0)\cdot\chi_{\lbrack
t_{k},\infty)}\in g(u\cdot\chi_{(-\infty,t_{k})}\oplus u(t_{k}-0)\cdot
\chi_{\lbrack t_{k},\infty)})
\]
follows from 100 (16) and from the fact that $f\subset g$.
\end{proof}

\begin{theorem}
We suppose that $(u,x)$ is a pseudo-fundamental mode of $f$ (a fundamental
mode of $f$, a fundamental mode of $f$ relative to $F$, a delay-insensitive
fundamental mode of $f$ relative to $F$), that $f$ is time invariant and let
$d\in\mathbf{R}$ so that $u\circ\tau^{d}\in S^{(m)}$. Then $(u\circ\tau
^{d},x\circ\tau^{d})$ is a pseudo-fundamental mode of $f$ (a fundamental mode
of $f$, a fundamental mode of $f$ relative to $F$, a delay-insensitive
fundamental mode of $f$ relative to $F$).
\end{theorem}

\begin{proof}
We suppose that $u$ is the constant function and from the time-invariance of
$f$ we have that $\forall x\in f(u),$ $x$ is the constant function (Corollary
43 a)). If some of 100 (13),...,(17) are true, then by the replacement of
$u,x$ with $u\circ\tau^{d}=u,$ $x\circ\tau^{d}=x$ the same statements are true.

We suppose now that $u$ is not constant, implying the existence of%
\[
t^{\prime}=\min\{t|u(t-0)\neq u(t)\}
\]
and the hypothesis $u\circ\tau^{d}\in S^{(m)}$ means that $t^{\prime}+d\geq0$.
The truth of some of the statements 100 (13),...,(17) implies the validity of
these statements after the replacement of $u,x,0\leq t_{0}<t_{1}<t_{2}<...$
with $u\circ\tau^{d},x\circ\tau^{d},0\leq t_{0}+d<t_{1}+d<t_{2}+d<...$ and we
have supposed without loss that $t_{0}=t^{\prime}$ (if $x$ is constant, this
statement is obvious and if $x$ is not constant, then
\[
t"=\min\{t|x(t-0)\neq x(t)\}
\]
exists and the non-anticipation -Definition 41- of $f$ gives $t^{\prime}\leq
t"$, see Corollary 43 b), so that $t_{0}=t^{\prime}$ is possible again).
\end{proof}

\begin{theorem}
Let the coordinatewise symmetrical Boolean function $F$ (Definition 70) and
the coordinatewise symmetrical system $f$ (Definition 71). If $(u,x)$ is a
pseudo-fundamental mode of $f$ (a fundamental mode of $f$, a fundamental mode
of $f$ relative to $F$, a delay-insensitive fundamental mode of $f$ relative
to $F$), then for all bijections $\sigma:\{1,...,m\}\rightarrow\{1,...,m\}$,
$(u_{\sigma},x)$ is a pseudo-fundamental mode of $f$ (a fundamental mode of
$f$, a fundamental mode of $f$ relative to $F$, a delay-insensitive
fundamental mode of $f$ relative to $F$).
\end{theorem}

\begin{proof}
From 100 (13),...,(17) and from the coordinatewise symmetry of $F$ and $f$ we
infer that%
\[
\forall t<t_{0},u_{\sigma}(t)=u_{\sigma}(t_{0}-0)
\]%
\[
\forall k\geq0,\forall t\in\lbrack t_{k},t_{k+1}),u_{\sigma}(t)=u_{\sigma
}(t_{k})
\]%
\[
\forall k\geq0,\forall t\in\lbrack t_{k},t_{k+1}),F(u_{\sigma}%
(t))=F(u(t))=F(u(t_{k}))=F(u_{\sigma}(t_{k}))
\]%
\begin{align*}
\forall k  &  \geq1,x\cdot\chi_{(-\infty,t_{k})}\oplus x(t_{k}-0)\cdot
\chi_{\lbrack t_{k},\infty)}\in f(u\cdot\chi_{(-\infty,t_{k})}\oplus
u(t_{k}-0)\cdot\chi_{\lbrack t_{k},\infty)})=\\
&  =f(u_{\sigma}\cdot\chi_{(-\infty,t_{k})}\oplus u_{\sigma}(t_{k}-0)\cdot
\chi_{\lbrack t_{k},\infty)})
\end{align*}%
\[
\forall k\geq1,x(t_{k}-0)=F(u(t_{k}-0))=F(u_{\sigma}(t_{k}-0))
\]
are fulfilled.
\end{proof}

\begin{theorem}
Let the rising-falling symmetrical function $F$ (Definition 79), the
rising-falling symmetrical system $f$ (Definition 80), $u\in S^{(m)}$ and
$x\in f(u)$. If $(u,x)$ is a pseudo-fundamental mode of $f$ (a fundamental
mode of $f$, a fundamental mode of $f$ relative to $F$, a delay-insensitive
fundamental mode of $f$ relative to $F$), then $(\overline{u},\overline{x})$
is a pseudo-fundamental mode of $f$ (a fundamental mode of $f$, a fundamental
mode of $f$ relative to $F$, a delay-insensitive fundamental mode of $f$
relative to $F$).
\end{theorem}

\begin{proof}
We infer from 100 (13),...,(17) and from the hypothesis of rising-falling
symmetry of $F,f$ that%
\[
\forall t<t_{0},\overline{u}(t)=\overline{u}(t_{0}-0)
\]%
\[
\forall k\geq0,\forall t\in\lbrack t_{k},t_{k+1}),\overline{u}(t)=\overline
{u}(t_{k})
\]%
\[
\forall k\geq0,\forall t\in\lbrack t_{k},t_{k+1}),F(\overline{u}%
(t))=\overline{F}(u(t))=\overline{F}(u(t_{k}))=F(\overline{u}(t_{k}))
\]%
\begin{align*}
\forall k  &  \geq1,\overline{x}\cdot\chi_{(-\infty,t_{k})}\oplus\overline
{x}(t_{k}-0)\cdot\chi_{\lbrack t_{k},\infty)}\in\overline{f}(u\cdot
\chi_{(-\infty,t_{k})}\oplus u(t_{k}-0)\cdot\chi_{\lbrack t_{k},\infty)})=\\
&  =f(\overline{u}\cdot\chi_{(-\infty,t_{k})}\oplus\overline{u}(t_{k}%
-0)\cdot\chi_{\lbrack t_{k},\infty)})
\end{align*}%
\[
\forall k\geq1,\overline{x}(t_{k}-0)=\overline{F}(u(t_{k}-0))=F(\overline
{u}(t_{k}-0))
\]
are true.\bigskip
\end{proof}

\bigskip{\Large 16. Generator function}

\begin{definition}
Let $\Phi:\mathbf{B}^{m}\times\mathbf{B}^{n}\rightarrow\mathbf{B}^{n},u\in
S^{(m)}$ and $x\in S^{(n)}$. We say that the state $x$ is generated by the
(generator) function $\Phi$ and the input (function) $u$ and that $\Phi,u$
generate (the state, the trajectory, the path) $x$ if the unbounded sequence
$0\leq t_{0}<t_{1}<t_{2}<...$ exists so that we have:%
\begin{equation}
u(t)=u(t_{0}-0)\cdot\chi_{(-\infty,t_{0})}(t)\oplus u(t_{0})\cdot\chi_{\lbrack
t_{0},t_{1})}(t)\oplus u(t_{1})\cdot\chi_{\lbrack t_{1},t_{2})}(t)\oplus...
\end{equation}%
\begin{equation}
x(t)=x(t_{0}-0)\cdot\chi_{(-\infty,t_{0})}(t)\oplus x(t_{0})\cdot\chi_{\lbrack
t_{0},t_{1})}(t)\oplus x(t_{1})\cdot\chi_{\lbrack t_{1},t_{2})}(t)\oplus...
\end{equation}%
\begin{equation}
\forall k\in\mathbf{N},\forall i\in\{1,...,n\},(x_{i}(t_{k+1})=x_{i}%
(t_{k})\quad or\quad x_{i}(t_{k+1})=\Phi_{i}(u(t_{k}),x(t_{k})))
\end{equation}%
\[
\{i|i\in\{1,...,n\},\exists k\in\mathbf{N},\exists a\in\mathbf{B}%
,a=x_{i}(t_{k})=x_{i}(t_{k+1})=...\quad and
\]%
\begin{equation}
and\quad\overline{a}=\Phi_{i}(u(t_{k}),x(t_{k}))=\Phi_{i}(u(t_{k+1}%
),x(t_{k+1}))=...\}=\emptyset
\end{equation}
\end{definition}

\begin{remark}
We interpret Definition 115 that formalizes in this context the unbounded
delay model from the asynchronous circuits theory.

a) For any $u,x$ an unbounded sequnce $(t_{k})$ like at (23), (24) exists.
These two equations fix such a sequence, that becomes the discrete time set.

b) Equations (25), (26) represent a restatement of Definition 2.10, items
$b),\,c)$ from [3] see also paragrapf 7 of that paper, by following an idea of
Anatoly Chebotarev. (25) states that for any (discrete) time moment $t_{k}$,
the new value of the coordinate $x_{i}$ is equal either with the old one, or
with $\Phi_{i}(u(t_{k}),x(t_{k}))$ (or with both). At (26) it is stated that
any computation of the 'next state' \ $\Phi_{i}(u(t_{k}),x(t_{k}))$ is
eventually made.

c) The common picture of all the trajectories that are generated by $\Phi$ and
$u$ was associated [2] with the propositional branching time temporal logic:
when $x_{i}(t_{k+1})=x_{i}(t_{k})$, respectively when $x_{i}(t_{k+1})=\Phi
_{i}(u(t_{k}),x(t_{k}))$, the 'proposition' $\ x$ runs in two different
branches of time.

d) Similarly with what happens at the fundamental mode, see Definition 103, if
$x$ is generated by $\Phi$ and $u$ and $0\leq t_{0}<t_{1}<t_{2}<...$ is an
unbounded sequence so that (23),...,(26) be true, we can call the sequnce
$(t_{k})$ compatible with $(u,x)$. Several sequences $(t_{k})$ exist that are
compatible with $(u,x)$, see for example the proof of Theorem 118.
\end{remark}

\begin{definition}
Let the state $x$ generated by $\Phi$ and $u$. The coordinate $i\in
\{1,...,n\}$ and the coordinate function $x_{i}$ are called excited or enabled
at the time instant $t$ if $x_{i}(t)\neq\Phi_{i}(u(t),x(t))$ and they are
called stable, or disabled at the time instant $t$ if $x_{i}(t)=\Phi
_{i}(u(t),x(t))$.

If $x(t)=\Phi(u(t),x(t))$ i.e. if all the coordinates are stable, we say that
the state $x$ is stable at the time instant $t$ and $(u(t),x(t))$ is called an
equilibrium point of $\Phi$.
\end{definition}

\begin{theorem}
We suppose that $x$ is generated by $\Phi$ and $u$. If it is stable at the
time instant $t^{\prime}$, then $\forall t\geq t^{\prime},x(t)=x(t^{\prime})$.
\end{theorem}

\begin{proof}
We suppose that \ some $k\in\mathbf{N}$ exists so that $x(t_{k})=\Phi
(u(t_{k}),x(t_{k}))$ (if the previous property is not true, then $x(t^{\prime
})=\Phi(u(t^{\prime}),x(t^{\prime}))$ is fulfilled for $t^{\prime}\notin
(t_{k})$; we reindex the elements of the set $t^{\prime}\cup(t_{k})$ and we
get an unbounded sequence $0\leq t_{0}^{^{\prime}}<t_{1}^{^{\prime}}%
<t_{2}^{^{\prime}}<...$ that makes (23),...,(26) from Definition 115 be
fulfilled and the property true). We have $x(t_{k})=x(t_{k+1})=...$
\end{proof}

\begin{notation}
The set of the states $x$ with $x(0-0)=x^{0}$ that are generated by $\Phi$ and
$u$ is noted with $L_{\Phi}(u,x^{0})$.
\end{notation}

\begin{remark}
$L_{\Phi}(u,x^{0})$ may be considered to be a $S^{(m)}\rightarrow P^{\ast
}(S^{(n)})$ $\ $function, i.e. an asynchronous system with the initial state
$x^{0}$.

On the other hand, we observe that for any $u$, some $x\in L_{\Phi}(u,x^{0})$
exists so that%
\[
u(t)=u(t_{0}-0)\cdot\chi_{(-\infty,t_{0})}(t)\oplus u(t_{0})\cdot\chi_{\lbrack
t_{0},t_{1})}(t)\oplus u(t_{1})\cdot\chi_{\lbrack t_{1},t_{2})}(t)\oplus...
\]%
\begin{align*}
x(t)  &  =x(t_{0}-0)\cdot\chi_{(-\infty,t_{0})}(t)\oplus x(t_{0}^{0})\cdot
\chi_{\lbrack t_{0}^{0},t_{0}^{1})}(t)\oplus...\oplus x(t_{0}^{p_{0}}%
)\cdot\chi_{\lbrack t_{0}^{p_{0}},t_{0}^{p_{0}+1})}(t)\oplus\\
&  \oplus x(t_{1}^{0})\cdot\chi_{\lbrack t_{1}^{0},t_{1}^{1})}(t)\oplus
...\oplus x(t_{1}^{p_{1}})\cdot\chi_{\lbrack t_{1}^{p_{1}},t_{1}^{p_{1}+1}%
)}(t)\oplus...
\end{align*}
where $x(t_{0}-0)=x(t_{0})=x^{0}$,%
\[
t_{0}=t_{0}^{0}<t_{0}^{1}<...<t_{0}^{p_{0}}<t_{0}^{p_{0}+1}=t_{1}=t_{1}%
^{0}<t_{1}^{1}<...<t_{1}^{p_{1}}<t_{1}^{p_{1}+1}=t_{2}=t_{2}^{0}<...
\]
$p_{0},p_{1},p_{2},...\in\mathbf{N}$ and%
\[
x(t_{k}^{j+1})=\Phi(u(t_{k}),x(t_{k}^{j})),j=\overline{0,p_{k}},k\in\mathbf{N}%
\]
It is interesting the situation when any $x\in L_{\Phi}(u,x^{0})$ is of this
form and the propositional branching time temporal logic becomes propositional
linear time temporal logic.
\end{remark}

\begin{definition}
We say that the system $f$ is generated by the (generator) function $\Phi$ if%
\[
\forall u,f(u)=\underset{x^{0}\in\phi(u)}{\bigcup}L_{\Phi}(u,x^{0})
\]
\end{definition}

\begin{example}
In the next four examples $m=n=1$, $\Phi:\mathbf{B}\times\mathbf{B}%
\rightarrow\mathbf{B},\mathbf{B}\times\mathbf{B}\ni(\lambda,\mu)\longmapsto
\Phi(\lambda,\mu)\in\mathbf{B}$ and $x^{0}$ is the initial state.

a) $\Phi(\lambda,\mu)=x^{1},x^{1}\in\mathbf{B}$ (the constant function)%
\[
L_{\Phi}(u,x^{0})=\{x|\exists t_{0}\geq0,x(t)=x^{0}\cdot\chi_{(-\infty,t_{0}%
)}(t)\oplus x^{1}\cdot\chi_{\lbrack t_{0},\infty)}(t)\}
\]
see also Theorem 123.

b) $\Phi(\lambda,\mu)=\lambda$ (the projection on the first coordinate)%
\begin{align*}
L_{\Phi}(u,x^{0})  &  =\{x|\text{ the unbounded sequence }0\leq t_{0}%
<t_{1}<t_{2}<...\text{ exists so that}\\
x(t)  &  =x^{0}\cdot\chi_{(-\infty,t_{0})}(t)\oplus u(t_{0})\cdot\chi_{\lbrack
t_{0},t_{1})}(t)\oplus u(t_{1})\cdot\chi_{\lbrack t_{1},t_{2})}(t)\oplus...\}
\end{align*}
Thus if $0\leq t_{0}^{^{\prime}}<t_{1}^{^{\prime}}<t_{2}^{^{\prime}}<...$ is
an unbounded sequence satisfying%
\[
u(t)=u(t_{0}^{\prime}-0)\cdot\chi_{(-\infty,t_{0}^{\prime})}(t)\oplus
u(t_{0}^{\prime})\cdot\chi_{\lbrack t_{0}^{\prime},t_{1}^{\prime})}(t)\oplus
u(t_{1}^{\prime})\cdot\chi_{\lbrack t_{1}^{\prime},t_{2}^{\prime})}%
(t)\oplus...
\]
and $0\leq t_{0}<t_{1}<t_{2}<...$ is a subsequence of $(t_{k}^{\prime})$, then
the state $x\in L_{\Phi}(u,x^{0})$ reproduces some of the successive values of
$u$ (infinitely many values). We remark that if $\underset{t\rightarrow\infty
}{\lim}$ $u(t)$ exists, then $\underset{t\rightarrow\infty}{\lim}$ $x(t)$
exists and $\underset{t\rightarrow\infty}{\lim}$ $u(t)=\underset
{t\rightarrow\infty}{\lim}$ $x(t)$.

c) $\Phi(\lambda,\mu)=\mu$ (the projection on the second coordinate)%
\[
L_{\Phi}(u,x^{0})=\{x^{0}\}
\]

d) $\Phi(\lambda,\mu)=\lambda\cdot\mu$%
\[
L_{\Phi}(u,x^{0})=\{x|\text{ the unbounded sequence }0\leq t_{0}<t_{1}%
<t_{2}<...\text{ exists so that}%
\]%
\[
x(t)=x^{0}\cdot\chi_{(-\infty,t_{0})}(t)\oplus x^{0}\cdot u(t_{0})\cdot
\chi_{\lbrack t_{0},t_{1})}(t)\oplus x^{0}\cdot u(t_{0})\cdot u(t_{1}%
)\cdot\chi_{\lbrack t_{1},t_{2})}(t)\oplus...\}
\]
Like at $b)$, $u(t_{0}),u(t_{1}),u(t_{2}),...$ are some of the successive
values talen by $u$.
\end{example}

\begin{theorem}
Let the function $\Phi$ and the initial state $x^{0}$. If $\exists x^{1}%
,\Phi=x^{1}$ (the constant function) then%
\[
L_{\Phi}(u,x^{0})=\{x|\forall i\in\{1,...,n\},\exists t_{i}\geq0,x_{i}%
(t)=x_{i}^{0}\cdot\chi_{(-\infty,t_{i})}(t)\oplus x_{i}^{1}\cdot\chi_{\lbrack
t_{i},\infty)}(t)\}
\]
\end{theorem}

\begin{proof}
In Definition 115, (25) shows for any $i$ that $x_{i}$ may switch from
$x_{i}^{0}$ to $x_{i}^{1}$ and (26) shows that if $x_{i}^{0}\neq x_{i}^{1}$
then some $t_{i}\geq0$ exists so that $x_{i}$ switches at $t_{i}$ from
$x_{i}^{0}$ to $x_{i}^{1}$.
\end{proof}

\begin{corollary}
If $f$ is generated by $\Phi=x^{1}$ then%
\[
\forall u,\forall x\in f(u),\exists x^{0}\in\phi(u),\forall i\in
\{1,...,n\},\exists t_{i}\geq0,\text{ }x_{i}(t)=x_{i}^{0}\cdot\chi
_{(-\infty,t_{i})}(t)\oplus x_{i}^{1}\cdot\chi_{\lbrack t_{i},\infty)}(t)
\]
\end{corollary}

\begin{theorem}
Let $f,\Phi,x^{0}$ and we suppose that%
\[
\forall u,f(u)=L_{\Phi}(u,x^{0})
\]

\begin{itemize}
\item[a)] If $\Gamma:\mathbf{B}^{m}\times\mathbf{B}^{n}\rightarrow
\mathbf{B}^{n}$ satisfies $\forall(\lambda_{1},...,\lambda_{m})\in
\mathbf{B}^{m},\forall(\mu_{1},...,\mu_{n})\in\mathbf{B}^{n},$%
\[
\Gamma(\lambda_{1},...,\lambda_{m},\mu_{1},...,\mu_{n})=\overline{\Phi
(\lambda_{1},...,\lambda_{m},\overline{\mu_{1}},...,\overline{\mu_{n}})}%
\]
then%
\[
\forall u,\overline{f}(u)=L_{\Gamma}(u,\overline{x^{0}})
\]

\item[b)] If $\Gamma:\mathbf{B}^{m+1}\times\mathbf{B}^{n}\rightarrow
\mathbf{B}^{n}$ satisfies $\forall(\lambda_{1},...,\lambda_{m+1})\in
\mathbf{B}^{m+1},\forall(\mu_{1},...,\mu_{n})\in\mathbf{B}^{n},$%
\[
\Gamma(\lambda_{1},...,\lambda_{m+1},\mu_{1},...,\mu_{n})=\Phi(\lambda
_{1},...,\lambda_{m},\mu_{1},...,\mu_{n})
\]
then%
\[
\forall(u_{1},...,u_{m+1})\in S^{(m+1)},f^{(m+1)}(u_{1},...,u_{m+1}%
)=L_{\Gamma}(u_{1},...,u_{m+1},x^{0})
\]

\item[c)] If $\Gamma:\mathbf{B}^{m}\times\mathbf{B}^{n}\rightarrow
\mathbf{B}^{n}$ satisfies for $i,j\in\{1,...,m\},i\neq j$: $\forall
(\lambda_{1},...,\lambda_{m})\in\mathbf{B}^{m},\forall(\mu_{1},...,\mu_{n}%
)\in\mathbf{B}^{n},$%
\[
\Gamma(\lambda_{1},...,\underset{i}{\lambda_{i}},...,\underset{j}{\lambda_{j}%
},...,\lambda_{m},\mu_{1},...,\mu_{n})=\Phi(\lambda_{1},...,\underset
{i}{\lambda_{i}},...,\underset{j}{\lambda_{i}},...,\lambda_{m},\mu_{1}%
,...,\mu_{n})
\]
then%
\[
\forall(u_{1},...,u_{m})\in S^{(m)},f_{i\rightarrow j}(u_{1},...,u_{m}%
)=L_{\Gamma}(u_{1},...,u_{m},x^{0})
\]

\item[d)] We suppose that $\Phi$ satisfies for some $i\in\{1,...,m\}:\forall
(\lambda_{1},...,\lambda_{m})\in\mathbf{B}^{m},\forall(\mu_{1},...,\mu_{n}%
)\in\mathbf{B}^{n},$%
\[
\Phi(\lambda_{1},...,\underset{i}{0},...,\lambda_{m},\mu_{1},...,\mu_{n}%
)=\Phi(\lambda_{1},...,\underset{i}{1},...,\lambda_{m},\mu_{1},...,\mu_{n})
\]
Then $f_{\widehat{u}_{i}}$ has sense and if $\Gamma:\mathbf{B}^{m-1}%
\times\mathbf{B}^{n}\rightarrow\mathbf{B}^{n}$ fulfills the condition
$\forall(\lambda_{1},...,\widehat{\lambda}_{i},...,\lambda_{m})\in
\mathbf{B}^{m-1},$ $\forall(\mu_{1},...,\mu_{n})\in\mathbf{B}^{n},$%
\[
\Gamma(\lambda_{1},...,\widehat{\lambda}_{i},...,\lambda_{m},\mu_{1}%
,...,\mu_{n})=\Phi(\lambda_{1},...,\underset{i}{0},...,\lambda_{m},\mu
_{1},...,\mu_{n})
\]
we have%
\[
\forall(u_{1},...,\widehat{u}_{i},...,u_{m})\in S^{(m-1)},f_{\widehat{u}_{i}%
}(u_{1},...,\widehat{u}_{i},...,u_{m})=L_{\Gamma}(u_{1},...,\widehat{u}%
_{i},...,u_{m},x^{0})
\]
\end{itemize}
\end{theorem}

\begin{proof}
At $a)$, if the equations (23),...,(26) from Definition 115 are fulfilled by
$u,x,\Phi$ then they are fulfilled by $u,\overline{x},\Gamma$ etc.
\end{proof}

\begin{remark}
A series of corollaries of Theorem 125 refers to the general case, when $f$ is
generated by $\Phi$, but it is not initialized. Another series of corollaries
of Theorem 125 follows from the supposition that $\Phi$ satisfies%
\[
\forall(\lambda_{1},...,\lambda_{m})\in\mathbf{B}^{m},\forall(\mu_{1}%
,...,\mu_{n})\in\mathbf{B}^{n},\Phi(\lambda_{1},...,\lambda_{m},\mu
_{1},...,\mu_{n})=\overline{\Phi(\lambda_{1},...,\lambda_{m},\overline{\mu
_{1}},...,\overline{\mu_{n}})}%
\]
at $a)$, (some examples of such functions for $m=1,$ $n=2$ are given by
$(\mu_{2},\lambda\cdot\mu_{1}\oplus\lambda\cdot\mu_{2}\oplus\lambda\oplus
\mu_{1}),$ $\quad(\mu_{2}\oplus1,\lambda\oplus\mu_{2}\oplus1),$ respectively
$(\lambda\oplus\mu_{1},\lambda\cdot\mu_{1}\oplus\lambda\cdot\mu_{2}\oplus
\mu_{1})$ ) and

$\forall(\lambda_{1},...,\lambda_{m})\in\mathbf{B}^{m},\forall(\mu_{1}%
,...,\mu_{n})\in\mathbf{B}^{n},\Phi(\lambda_{1},...,\underset{i}{\lambda_{i}%
},...,\underset{j}{\lambda_{j}},...,\lambda_{m},\mu_{1},...,\mu_{n}%
)=\Phi(\lambda_{1},...,\underset{i}{\lambda_{j}},...,\underset{j}{\lambda_{i}%
},...,\lambda_{m},\mu_{1},...,\mu_{n})$

respectively

$\forall(\lambda_{1},...,\lambda_{m})\in\mathbf{B}^{m},\forall(\mu_{1}%
,...,\mu_{n})\in\mathbf{B}^{n},\Phi(\lambda_{1},...,\underset{i}{\lambda_{i}%
},...,\underset{j}{0},...,\lambda_{m},\mu_{1},...,\mu_{n})=\Phi(\lambda
_{1},...,\underset{i}{\lambda_{i}},...,\underset{j}{1},...,\lambda_{m},\mu
_{1},...,\mu_{n})$

at $c)$.

On the other hand systems exist that are not generated by any function, for
example those from Example 3 (1),(3),(4) that are characterized by the
parameters $d\geq0,\delta_{r}\geq0,\delta_{f}\geq0$ are in this situation.

The problem of the generator functions leaves open a lot of questions, from
the generation of the intersection and the reunion of the systems, to the
connections with other topics from our work, such as the parallel connection
and the serial connection, the symmetry in both variants and the stability.
\end{remark}

\bigskip

{\Large Appendix. Details related with Remark 10}

\bigskip

With $u^{1}\in S^{(m_{1})},...,u^{p}\in S^{(m_{p})}$ we form the functions
$(u^{1},...,u^{p}):\mathbf{R}^{p}\rightarrow\mathbf{B}^{m_{1}}\times
...\times\mathbf{B}^{m_{p}},$%
\[
\forall(t_{1},...,t_{p})\in\mathbf{R}^{p},(u^{1},...,u^{p})(t_{1}%
,...,t_{p})=(u^{1}(t_{1}),...,u^{p}(t_{p}))
\]
$(u^{1},...,u^{p})\in S^{(m_{1})}\times...\times S^{(m_{p})}$ and respectively
$u^{1}\intercal...\intercal u^{p}:\mathbf{R}\rightarrow\mathbf{B}%
^{m_{1}+...+m_{p}},$%
\[
\forall t\in\mathbf{R},(u^{1}\intercal...\intercal u^{p})(t)=(u_{1}%
^{1}(t),...,u_{m_{1}}^{1}(t),...,u_{1}^{p}(t),...,u_{m_{p}}^{p}(t))
\]
$u^{1}\intercal...\intercal u^{p}\in S^{(m_{1}+...+m_{p})}.$ A bijection
$\pi:S^{(m_{1})}\times...\times S^{(m_{p})}\rightarrow S^{(m_{1}+...+m_{p})}$
exists,%
\[
\forall(u^{1},...,u^{p})\in S^{(m_{1})}\times...\times S^{(m_{p})},\pi
(u^{1},...,u^{p})=u^{1}\intercal...\intercal u^{p}%
\]
allowing us to identify $S^{(m_{1})}\times...\times S^{(m_{p})}$ with
$S^{(m_{1}+...+m_{p})}$.

We form two sets with $X_{1}\in P^{\ast}(S^{(n_{1})}),...,X_{p}\in P^{\ast
}(S^{(n_{p})})$: $\,(X_{1},...,X_{p})\in P^{\ast}(S^{(n_{1})})\times...\times
P^{\ast}(S^{(n_{p})})$ and respectively $X_{1}\intercal...\intercal X_{p}\in
P^{\ast}(S^{(n_{p}+...+n_{p})})$ that is defined this way%
\[
X_{1}\intercal...\intercal X_{p}=\{x^{1}\intercal...\intercal x^{p}|x^{1}\in
X_{1},...,x^{p}\in X_{p}\}
\]
We have a bijection $\Pi:P^{\ast}(S^{(n_{1})})\times...\times P^{\ast
}(S^{(n_{p})})\rightarrow P^{\ast}(S^{(n_{p}+...+n_{p})}),$%
\[
\forall(X_{1},...,X_{p})\in P^{\ast}(S^{(n_{1})})\times...\times P^{\ast
}(S^{(n_{p})}),\Pi(X_{1},...,X_{p})=X_{1}\intercal...\intercal X_{p}%
\]
that allows us to identify the sets $P^{\ast}(S^{(n_{1})})\times...\times
P^{\ast}(S^{(n_{p})})$ and $P^{\ast}(S^{(n_{p}+...+n_{p})}).$

With the functions $f^{i}:S^{(m_{i})}\rightarrow P^{\ast}(S^{(n_{i}%
)}),i=\overline{1,p}$ we form two functions $(f^{1},...,f^{p}):S^{(m_{1}%
)}\times...\times S^{(m_{p})}\rightarrow P^{\ast}(S^{(n_{1})})\times...\times
P^{\ast}(S^{(n_{p})}),$%
\[
\forall(u^{1},...,u^{p})\in S^{(m_{1})}\times...\times S^{(m_{p})}%
,(f^{1},...,f^{p})(u^{1},...,u^{p})=(f^{1}(u^{1}),...,f^{p}(u^{p}))
\]
and respectively $f^{1}\intercal...\intercal f^{p}:S^{(m_{1}+...+m_{p}%
)}\rightarrow P^{\ast}(S^{(n_{1}+...+n_{p})}),$%
\[
\forall(u^{1}\intercal...\intercal u^{p})\in S^{(m_{1}+...+m_{p})}%
,(f^{1}\intercal...\intercal f^{p})(u^{1}\intercal...\intercal u^{p}%
)=f^{1}(u^{1})\intercal...\intercal f^{p}(u^{p})
\]
The commutativity of the diagram%
\[%
\begin{array}
[c]{ccc}%
\begin{array}
[c]{c}%
\\
S^{(m_{1})}\times...\times S^{(m_{p})}%
\end{array}
& \underrightarrow{\,\,(f^{1},...,f^{p})\,\,} &
\begin{array}
[c]{c}%
\\
P^{\ast}(S^{(n_{1})})\times...\times P^{\ast}(S^{(n_{p})})
\end{array}
\\
\pi\downarrow &  & \downarrow\Pi\\%
\begin{array}
[c]{c}%
\\
S^{(m_{1}+...+m_{p})}%
\end{array}
& \underrightarrow{\,f^{1}\intercal...\intercal f^{p}\,} &
\begin{array}
[c]{c}%
\\
P^{\ast}(S^{(n_{1}+...+n_{p})})
\end{array}
\end{array}
\]
makes us identify the functions $(f^{1},...,f^{p})$ and $f^{1}\intercal
...\intercal f^{p}$.

\end{document}